\documentclass[onecolumn]{emulateapj}
\shorttitle{QSO Mass Functions}
\shortauthors{Kelly et al.}

\begin{document}
  
  \title{Constraints on Black Hole Growth, Quasar Lifetimes, and
    Eddington Ratio Distributions from the SDSS Broad Line Quasar
    Black Hole Mass Function}

  \author{Brandon C. Kelly\altaffilmark{1,2,3}, Marianne
    Vestergaard\altaffilmark{4,5}, Xiaohui Fan\altaffilmark{5,6}, Philip
    Hopkins\altaffilmark{7}, Lars Hernquist\altaffilmark{3}, Aneta
    Siemiginowska\altaffilmark{3}}

  \altaffiltext{1}{bckelly@cfa.harvard.edu}
  \altaffiltext{2}{Hubble Fellow}
  \altaffiltext{3}{Harvard-Smithsonian Center for Astrophysics, 
      60 Garden St, Cambridge, MA 02138}
  \altaffiltext{4}{Freja Fellow, Dark Cosmology Centre, The Niels Bohr
  Institute, University of Copenhagen}
  \altaffiltext{5}{Department of Astronomy, University of Arizona,
    Tucson, AZ 85721} 
  \altaffiltext{6}{Max Planck Institute for Astronomy, Heidelberg, Germany}
  \altaffiltext{7}{Miller Fellow, Department of Astronomy, University of California,
    Berkeley, CA}

  \begin{abstract}

    We present an estimate of the black hole mass function
    (BHMF) of broad line quasars (BLQSOs) that self-consistently corrects
    for incompleteness and the statistical uncertainty in the mass
    estimates, based on a sample of 9886 quasars at $1 < z < 4.5$ drawn
    from the Sloan Digital Sky Survey. We find evidence for `cosmic
    downsizing' of black holes in BLQSOs, where the peak in their number
    density shifts to higher redshift with increasing black hole mass. The
    cosmic mass density for black holes seen as BLQSOs peaks at $z \sim
    2$. We estimate the completeness of the SDSS as a function of black
    hole mass and Eddington ratio, and find that at $z > 1$ it is highly incomplete
    at $M_{BH} \lesssim 10^9 M_{\odot}$ and $L / L_{Edd} \lesssim 0.5$. We
    estimate a lower limit on the lifetime of a single BLQSO phase to be
    $t_{BL} > 150 \pm 15$ Myr for black holes at $z=1$ with a mass of
    $M_{BH} = 10^9 M_{\odot}$, and we constrain the maximum mass of a
    black hole in a BLQSO to be $\sim 3 \times 10^{10} M_{\odot}$. Our
    estimated distribution of BLQSO Eddington ratios peaks at $L / L_{Edd}
    \sim 0.05$ and has a dispersion of $\sim 0.4$ dex, implying that most
    BLQSOs are not radiating at or near the Eddington limit; however
    the location of the peak is subject to considerable uncertainty. The steep increase in number
    density of BLQSOs toward lower Eddington ratios is expected if the
    BLQSO accretion rate monotonically decays with time. Furthermore, our
    estimated lifetime and Eddington ratio distributions imply that the
    majority of the most massive black holes spend a significant amount of
    time growing in an earlier obscured phase, a conclusion which is
    independent of the unknown obscured fraction. These results are
    consistent with models for self-regulated black hole growth, at least
    for massive systems at $z > 1$, where the BLQSO phase occurs at the
    end of a fueling event when black hole feedback unbinds the accreting
    gas, halting the accretion flow.

  \end{abstract}
  
  \keywords{galaxies: active --- galaxies: mass function --- galaxies:
    statistics}
  
  \section{INTRODUCTION}

  \label{s-intro}

  Understanding how and when supermassive black holes (SMBHs) grow is
  currently of central importance in extragalactic
  astronomy. Observations have established correlations between SMBH
  mass and host galaxy spheroidal properties, such as luminosity
  \citep[e.g.,][]{korm95,mclure01,mclure02}, stellar velocity
  dispersion \citep[$M_{BH}$--$\sigma$ relationship, e.g.,][]{gebh00a,
    merr01, trem02}, concentration or Sersic index
  \citep[e.g.,][]{graham01,graham07}, bulge mass
  \citep[e.g.,][]{mag98,marc03,haring04}, and binding energy
  \citep[e.g.,][]{aller07,fundplane}. These correlations imply that
  the evolution of spheroidal galaxies and the growth of SMBHs are
  intricately tied together, where black holes grow by accreting gas,
  possibly fueled by a major merger of two gas-rich galaxies, until
  feedback energy from the SMBH expels gas and shuts off the accretion
  process
  \citep[e.g.,][]{silk98,fabian99,begel05,murray05,hmt09}. This
  `self-regulated' growth of black holes has been successfully applied
  in smoothed particle hydrodynamics simulations
  \citep{dimatt05,spring05,johan09}, and has motivated numerous models linking
  the SMBH growth, the quasar phase, and galaxy evolution
  \citep[e.g.,][and references
  therein]{haehn98,kauff00,haehn00,wyithe03,vol03,catt05,dimatteo08,
    somer08,hopkins_cosmo,croton09,sijacki09,booth09,shen09}. Within this framework,
  the broad line quasar\footnote[8]{Throughout this work we will use
    the terms quasar and AGN to refer generically to broad line
    AGNs. No luminosity difference between the two is assumed, but
    they are assumed to have broad emission lines along our line of sight.}  phase persists
  after feedback energy from the black hole `blows' the gas away
  \citep[e.g.,][]{quasar_evol,hopkins_long}. The broad line quasar
  phase is expected to persist until the accretion rate drops low enough to switch to
  a radiatively inefficient accretion flow
  \citep[e.g.,][]{churazov05,hhqh09}.

  While major-mergers of gas-rich galaxies may fuel quasars at high
  redshift, and grow the most massive SMBHs, alternative fueling
  mechanisms are likely at lower redshift and fainter
  luminosities. Mergers alone do not appear to be sufficient to
  reproduce the number of X-ray faint AGN \citep[e.g.,][]{marulli07}, and
  accretion of ambient gas \citep[e.g.,][]{ciotti01,stochacc}, may
  fuel these fainter, lower $M_{BH}$ AGN at lower $z$, resulting in an
  alternative growth mechanism for these SMBHs. Indeed, many AGN are
  observed to live in late-type galaxies out to $z \approx 1$
  \citep[e.g.,][]{guyon06,gabor09}, and the X-ray luminosity function
  of AGN hosted by late-type galaxies suggests that fueling by minor
  interactions or internal instabilities represents a non-negligible
  contribution to the accretion history of the Universe
  \citep{georg09}. Furthermore, feedback from the SMBH may continue to
  affect its environment long after its growth through so-called
  `radio mode' feedback
  \citep{croton06,bower06,sijacki07}. Observations qualitatitively
  support a model where the fueling mechanism for black hole growth
  depends on the mass of the host dark matter halo, but regardless of
  the fueling mechanism, black hole feedback and accretion follow a
  similar evolutionary path \citep[e.g.,][]{hickox09,const09}.

  Observationally, a significant amount of recent work has utilized
  the argument of \citet{soltan82} to indirectly map the growth of all
  SMBHs
  \citep[e.g.,][]{sal99,yu02,shank04,shank09,marconi04,yu04,hopkins07,merloni08}. Work
  along this line has used the correlations between $M_{BH}$ and host
  galaxy spheroidal properties to infer the local distribution of
  $M_{BH}$ for inactive black holes, which are assumed to be the
  relics of past AGN activity. The distribution of $M_{BH}$ as a
  function of redshift is then estimated by stepping backward from the
  local distribution of $M_{BH}$, employing a continuity equation
  describing the `flow' of black hole number density through bins in
  $M_{BH}$ \citep[e.g.,][]{small92}. The quasar luminosity function is
  used as a constraint on the rate of change in the SMBH mass density,
  because it traces the accretion of matter onto black holes, modulo
  the accretion efficiency and the bolometric correction. From this
  work, it has generally been inferred that black hole growth is
  dominated by periods of near Eddington accretion, with the most
  massive SMBHs growing first, and that many SMBHs have non-zero
  spin.

  An alternative to the technique of \citet{soltan82} for estimating
  the SMBH mass function has been used by \citet{siem97} and
  \citet{hatz01}. These authors used a model for thermal-viscous
  accretion disk instabilities \citep{siem96} to calculate the
  expected distribution of luminosity at a given black hole
  mass. Based on this calculated distribution, they use the quasar
  luminosity function to constrain the quasar black hole mass
  function. \citet{siem97} found evidence for black hole `downsizing',
  with the peak of the quasar mass function shifting toward lower
  masses at lower redshift.

  Correlations between the SMBH mass, width of the broad emission
  lines, and luminosity of the quasar continuum have made it possible
  to estimate $M_{BH}$ for broad line quasars (BLQSOs)
  \citep[e.g.,][]{vest06,kelly07b}, 
  albeit with considerable statistical uncertainty of $\sim 0.4$ dex
  and various systematic effects
  \citep[e.g.,][]{krolik01,vest06,greene06,marconi08,denney09}. This
  offers an 
  alternative to estimating SMBH mass functions based on the
  \citet{soltan82} argument, because the mass function may be
  estimated directly at all redshifts, and because the distribution of
  quasar emission line widths provides an additional observational constraint
  on the mass function. Estimates of
  $M_{BH}$ obtained from the broad emission lines have been used to
  estimate the distribution of quasar black hole masses and Eddington
  ratios at a variety of redshifts
  \citep[e.g.,][]{mclure04,vest04,koll06,netztrak07,shen08,fine08,kelly08,trump09,labita09a}.

  The BLQSO black hole mass function (BHMF) maps
  the comoving number density and evolution of active supermassive
  black holes contained within broad line AGN, and is therefore a
  complete census of the population of these SMBHs over cosmic
  time. The BLQSO BHMF is important for a number of reasons, including
  the following:
  \begin{itemize}
  \item
    At high redshift SMBHs with masses $M_{BH} \sim 10^9$ are already in place by $z \sim
    6$ \citep[e.g.,][]{fan01b,jiang07}, and therefore the high-$z$
    BLQSO BHMF places important constraints on the formation and
    growth of seed SMBHs \citep[e.g.,][]{vol03,vol08,lodato07}. 
  \item
    If, after a fueling event, the growth of the SMBH
    persists until it becomes massive enough such that feedback energy
    begins to unbind the gas, the active SMBH will be seen
    as a BLQSO shortly before entering quiescence, and its fractional
    mass growth will not be significant during this time period
    \citep{stochacc}. The BLQSO BHMF 
    thus (1) is related to the distribution of spheroidal binding energies
    in the central regions of the galaxy \citep{hop07a,fundplane,younger08},
    and (2) gives the nearly instantaneous increase in the AGN relic SMBH mass
    density. 
  \item
    Because mass is a fundamental physical quantity of SMBHs, we
    can use the BLQSO BHMF to estimate the duty cycle for broad line quasar activity as a
    function of mass by comparing the number density of all SMBHs with
    those seen as BLQSOs. The duty cycle can then be converted into an
    estimate of the lifetime of BLQSO activity, which is of
    significant importance for understanding the origin of BLQSO
    activity. This cannot be done using the quasar luminosity function.
  \item
    The distribution of luminosities at a given
    BLQSO SMBH mass depends on the quasar lightcurve
    \citep{yu04,stochacc,yu08,notlightbulbs}, which is a function of
    both evolution in the rate at which fuel is supplied to the
    accretion disk, and the time-dependent behavior of the accretion
    disk \citep{siem97,hatz01}. Thus, understanding the distribution
    of $L$ at a given $M_{BH}$, or alternatively the distribution of
    Eddington ratio, gives insight into the BLQSO fueling mechanism
    and accretion physics.
  \end{itemize}
  The BLQSO BHMF therefore provides an important observational
  constraint on models of SMBH growth, the onset and duration of
  quasar activity, quasar feedback, and galaxy evolution.

  There have been several estimates of the BHMF of BLQSOs calculated directly
  from the mass estimates derived from the broad emission lines
  \citep[][hereafter KVF09]{wang06,greene07,vest08,vest09,kvf09}. In particular, \citet{vest09}
  found evidence for cosmic `downsizing' of black hole mass, in that the
  most massive SMBHs are more common at high redshift, consistent with
  previous work on mapping black hole growth
  \citep[e.g.,][]{marconi04,merloni04,shank04,merloni08},
  conclusions based on the quasar luminosity function
  \citep[e.g.,][]{steffen03,ueda03,croom04,lafranca05,has05,hopkins07,silv08}
  and the local active BHMF \citep{heckman04}. However,
  a major concern with previous estimates of the BLQSO BHMF is
  uncorrected incompleteness and the additional broadening caused by
  the statistical errors in mass estimates
  \citep[][KVF09]{kelly07b,shen08}. Because the massive end of the BLQSO 
  BHMF falls off with increasing $M_{BH}$, the intrinsic uncertainty
  scatters more sources into higher $M_{BH}$ bins than lower ones,
  potentially having a significant effect on the estimated number
  density of the most massive SMBHs. Furthermore, even if a sample is
  complete in luminosity, it is not necessarily complete in $M_{BH}$,
  and the completeness in $M_{BH}$ depends on the unknown Eddington
  ratio distribution. Motivated by these issues, and the
  fact that the importance of the BLQSO BHMF demands that its
  determination be statistically rigorous, KVF09 developed a Bayesian
  statistical technique for estimating the BLQSO BHMF that
  self-consistently corrects for the incompleteness in $M_{BH}$ and
  the statistical uncertainties in the estimates of $M_{BH}$. In this
  work we apply the technique of KVF09 to the SDSS BLQSO sample of
  \citet{vest08} in order to estimate the black hole mass
  function of SMBHs that reside in BLQSOs, and discuss the
  implications for SMBH growth, quasar lifetimes, and Eddington ratio
  distributions.

  \section{THE DATA}

  \label{s-data}

  Our sample is drawn from the Sloan Digital Sky Survey (SDSS) Data
  Release 3 (DR3) quasar sample as presented by \citet{dr3lumfunc} and
  \citet{vest08}. \citet{dr3lumfunc} used 15,180 quasars from the SDSS
  DR3 to determine the quasar optical luminosity function over $0.3 <
  z < 5$. \citet{vest08} obtained black hole estimates for 14,434 of
  the quasars presented in \citet{dr3lumfunc} using scaling
  relationships between the width of the broad emission lines,
  continuum luminosity, and black hole mass. The details of the sample
  and fitting process are described in \citet{dr3lumfunc} and
  \citet{vest08}. For completeness, we briefly review the spectral
  modeling used by \citet{vest08} to extract the relevant quantities.

  \citet{vest08} modeled the observed quasar spectra using a power-law
  continuum, an optical-UV iron line spectrum based on I Zw I
  \citep{uvfe,optfe}, a Balmer continuum, and host galaxy template
  \citep{bruz03} for objects at $z < 0.5$. The continuum luminosities
  used for the mass estimates are based on the power-law continuum
  fits. The emission lines used for the mass estimates in this work are Mg II and C
  IV, and were modeled using multiple Gaussian
  functions so to reproduce the line profile. Contributions from the
  narrow emission line region were
  subtracted from Mg II when appropriate, and line profiles with strong
  absorption were ignored. The end result of this analysis was a set
  of emission line FWHM and continuum luminosities, from which black
  hole mass estimates were calculated.

  We have performed a few additional cuts to the sample presented by
  \citet{vest08} before obtaining our final sample. First, we remove
  the sources located at $z < 1$. We do this because the quasar
  sample contains a significant number of extended sources below $z <
  1$, and therefore their $i$-band magnitudes sometimes contain a
  significant contribution from the host galaxy (see
    the discussion in \citet{dr3lumfunc}). This reduces the
  effective flux limit at $z < 1$, creating an artificial second peak in
  the redshift distribution. While we could attempt to empirically correct the
  selection function to account for this, we find it easier to simply
  limit our analysis to $1 < z < 4.5$. Finally, in order to make our
  analysis more robust against uncertainty in the selection function,
  we omit any sources for which the value of the selection function of
  \citet{dr3lumfunc} is less than $0.01$, and 
  force all values of the selection function to be zero that are $<
  0.01$. After making these cuts, we were left with a sample of 9886
  quasars.

  \section{THE STATISTICAL MODEL AND DATA ANALYSIS}

  \label{s-statmod}

  We use an expanded version of the statistical model outlined in KVF09 to estimate the BLQSO
  BHMF. For completeness, we describe the important aspects 
  of the technique developed by KVF09, and the reader is referred to
  KVF09 for further details regarding the technique and its
  derivation. Qualitatively, the technique of KVF09 assumes
  parameteric forms for the BHMF and the distribution of luminosities
  at a given $M_{BH}$.  The average value of the broad line mass
  estimates at a given $M_{BH}$ is fixed
  to be consistent with the scaling relationships reported by
  \citet{vest06} and \citet{vest09}; we assume that these scaling
  relationships produce unbiased estimates of $M_{BH}$. The BHMF and distribution of $L$ at a given
  $M_{BH}$, in combination with a selection function, imply an
  observed distribution of $z, L,$ and FWHM. The technique of KVF09
  attempts to recover the BHMF and distribution of $L$ at a given
  $M_{BH}$ by matching the observed distribution of $z,L,$ and
  FWHM (or, equivalently, the mass estimates) that is implied by the model to the actual observed
  distribution. The posterior probability 
  distribution is used to quantify how well the implied distributions
  match the observed distributions. As a result, the technique of KVF09
  estimates the probability distribution of the BHMF, and of the
  distribution of $L$ at a given $M_{BH}$, given the observed data
  set.

  \subsection{Model for the Joint Distribution of $M_{BH}, L, {\rm
      FWHM},$ and $z$}

  \label{s-statmod2}
  We model the BLQSO BHMF as a mixture of $K$ log-normal
  distributions:
  \begin{equation}
    \phi(\log M_{BH},\log z) = N \left(\frac{dV}{dz}\right)^{-1} 
    \sum_{k=1}^{K} \frac{\pi_k}{2 \pi
      |\Sigma_k|^{1/2}} \exp \left[ -\frac{1}{2} ({\bf y} - \mu_k)^T
      \Sigma_k^{-1} ({\bf y} - \mu_k) \right],
    \label{eq-mixmod}
  \end{equation}
  where $\sum_{k=1}^{K} \pi_k = 1$. In this work we choose $K = 4$,
  based on our previous experience in working with simulated data sets
  (KVF09), and because we did not notice a significant difference in
  the results obtained using larger values of $K$. Here, $N$ is the
  total number of BLQSOs in the 
  Universe that could be seen by an observer on Earth at the time of
  the survey, ${\bf y} = (\log M_{BH},\log z)$, $\mu_k$ is the
  2-element mean vector for the $k^{\rm th}$ Gaussian functions,
  $\Sigma_k$ is the $2 \times 2$ covariance matrix for the $k^{\rm
    th}$ Gaussian function, and $x^T$ denotes the transpose of $x$. In
  addition, we denote $\pi = (\pi_1, \ldots, \pi_K), \mu = (\mu_1,
  \ldots, \mu_K)$, and $\Sigma = (\Sigma_1, \ldots, \Sigma_K)$. The
  variance in $\log M_{BH}$ for Gaussian function $k$ is
  $\sigma^2_{m,k} = \Sigma_{11,k}$, the variance in $\log z$ for
  Gaussian function $k$ is $\sigma^2_{z,k} = \Sigma_{22,k}$, and the
  covariance between $\log M_{BH}$ and $\log z$ for Gaussian function
  $k$ is $\sigma_{mz,k} = \Sigma_{12,k}$. The parameters for the mass
  function are $N, \pi, \mu,$ and $\Sigma$.

  The distribution of luminosity density at a given black hole mass
  and wavelength $\lambda$ is assumed to also follow a mixture of $J$
  log-normal distributions:
  \begin{equation}
     p(\log L_{\lambda}|M_{BH}) = \sum_{j=1}^{J} \frac{\gamma_j}{\sqrt{2 \pi \sigma_{l,j}^2}} \exp \left[
      -\frac{1}{2} \left( \frac{\log \lambda L_{\lambda} - \alpha_{0,j} -
          \alpha_{m,j} (\log M_{BH} - 9)}{\sigma_{l,j}}
      \right)^2 \right].
    \label{eq-problm}
  \end{equation}
  This represents an extension of the model of KVF09, which only used
  $J = 1$ log-normal distribution. We made this extension to
  incorporate additional flexiblity in $p(L_{\lambda}|M_{BH})$, ensuring that 
  the luminosity distribution, and therefore the
  black hole mass incompleteness correction, is robust to the particular
  parameteric form. For each log-normal distribution, the parameters
  for the distribution of $L_{\lambda}$ at a given 
  $M_{BH}$ are $\gamma_j$, $\alpha_{0,j}, \alpha_{m,j},$ and
  $\sigma_{l,j}$. We used $J = 3$ log-normal distributions, as we did
  not notice a signficant change in the estimated values of
  $p(L_{\lambda}|M_{BH})$ when using $J \geq 3$. We further assess the
  robustness of our assumed form for $p(L_{\lambda}|M_{BH})$ in
  \S~\ref{s-incorrect_lcurve}, and show that our results should not be
  signficantly altered if the true form of $p(L_{\lambda}|M_{BH})$ is a power-law,
  as might be expected from some physical models for BLQSO
  lightcurves.

  In our analysis we
  use the luminosity density at 1350\AA\ in order to minimize bias at
  the highest redshifts introduced from extrapolating the power-law
  continuum beyond the spectral window. At $z \gtrsim 1.8$ the rest
  frame $\lambda = 1350$\AA\ falls within the observed spectral window
  for the SDSS sources used in this work, while a smaller redshift
  window is available when using the luminosity densities calculated
  at other wavelengths.  Furthermore, the $z \sim 1$ distributions of
  bolometric luminosity did not exhibit any significant difference
  when using the luminosity density at 1350\AA, as compared to that
  calculated using the luminosity density at $\lambda > 1350$\AA,
  suggesting that significant biases at lower $z$ are not introduced by
  extrapolating the power-law continuum.

  We can connect the parameters in Equation (\ref{eq-problm}) to the
  distribution of Eddington ratio and bolometric correction. The
  monochromatic luminosity is related to the Eddington luminosity
  ratio $\Gamma_{Edd}$\footnote[9]{We will occasionally use
    $\Gamma_{Edd}$ to denote the Eddington ratio, instead of $L /
    L_{Edd}$. We do this in certain instances where it is more
    notationally convenient to do so.} and bolometric correction $C_{\lambda}$ as
  \begin{equation}
    \lambda L_{\lambda} = 1.3 \times 10^{38} \frac{\Gamma_{Edd}}{C_{\lambda}}
    \frac{M_{BH}}{M_{\odot}} \ \ [{\rm erg\ s^{-1}}].
      \label{eq-mlrel}
  \end{equation}
  Therefore, Equation (\ref{eq-problm}) implies that for the $j^{\rm
    th}$ log-normal distribution we are assuming
  that on average $\log (\Gamma_{Edd} / C_{\lambda}) = \alpha_{0,j} - 47.11 +
  (\alpha_{m,j} - 1) \log M_{BH}$, with a Gaussian scatter about this mean
  value of standard deviation $\sigma_{l,j}$. We do not make any formal
  attempt to prohibit Equation (\ref{eq-problm}) from allowing values
  of $L / L_{Edd} > 1$, as this would require us to make an
  assumption about the bolometeric correction. However, as we will
  show in \S~\ref{s-eddrat}, our estimate for $p(L_{\lambda}|M_{BH})$
  implies only a negligible fraction of BLQSOs with $L / L_{Edd} >
  1$, assuming a constant bolometric correction of $C_{1350} = 4.3$
  \citep{vest09}.

  The distribution of emission line widths at a given luminosity
  density and black hole mass is modeled as a log-normal distribution:
  \begin{eqnarray}
    \lefteqn{p(\log {\rm FWHM}_{BL}|L_{BL},M_{BH}) =} \nonumber \\ 
    & & \frac{1}{\sqrt{2 \pi (\sigma_{BL}^2 + \sigma^2_{\rm FWHM})}} \exp \left\{
      -\frac{1}{2} \frac{(\log {\rm FWHM}_{BL} - \beta^{BL}_0 + 
          1/4 \log L_{BL} - 1/2 \log
          M_{BH})^2}{\sigma^2_{BL} + \sigma^2_{\rm FWHM}} \right\}.
    \label{eq-probvlm}
  \end{eqnarray}
  Here, ${\rm FWHM}_{BL}$ is the line width for a particular broad emission
  line, $L_{BL}$ is the luminosity density used as an estimate for the
  broad line region size for a particular broad emission line,
  $\sigma_{\rm FWHM}$ is the measured uncertainty in $FWHM$ due to measurement
  error, and
  $\beta_0^{BL}$ and $\sigma_{BL}$ are the parameters for Equation
  (\ref{eq-probvlm}) for a particular broad emission line. We do not
  make any attempt to correct for radiation pressure on the broad
  emission line clouds \citep{marconi08}, as its importance and
  effects are currently poorly understood \citep[e.g., for a
  discussion see][]{vest09}.

  Equation (\ref{eq-probvlm}) implies that on average ${\rm FWHM}
  \propto M^{1/2}_{BH} / L_{BL}^{1/4}$, or equivalently $M_{BH}
  \propto L^{1/2}_{BL} {\rm FWHM}^2_{BL}$. Therefore, we can use the
  results obtained for the broad emission line scaling estimates of
  $M_{BH}$ to fix $\beta_0$. We use the mass scaling relationship for
  C IV that is presented by \citet{vest06}, and the
  relationship for Mg II that is presented by
  \citet{vest09}. As noted in KVF09, these scaling
  relationships imply $\beta^{\rm
    MgII}_0 = 10.61$ and $\beta^{C IV}_0 = 11.33$, and we fix
  $\beta_0$ to these values.

  In addition, the
  dispersion in ${\rm FWHM}_{BL}$ at a given $L_{BL}$ and $M_{BH}$ can be
  related to the scatter in the mass estimates based on the scaling
  relationships. \citet{vest06} find the statistical uncertainty in
  the broad line mass estimates to be 0.36 dex for
  C IV. Therefore, because ${\rm FWHM} \propto
  M_{BH}^{1/2}$, $\sigma_{BL}({\rm CIV}) \approx 0.18$ dex. Likewise,
  the intrinsic uncertainty in the broad line mass estimate for Mg II
  is found to be $\sim 0.4$ dex (Vestergaard et al., in preparation), and
  therefore $\sigma_{BL}({\rm MgII}) \approx 0.2$ dex. However, there
  have been indications from previous analysis of flux limited surveys, which probe 
  higher redshifts and a narrower range in luminosity than that exhibited by the objects with
  reverberation mapping data, that the intrinsic scatter in the mass
  estimates may be smaller than $\approx 0.4$ dex
  \citep{koll06,shen08,fine08,stein10a,stein10b}. This
  may be caused by correlated scatter in the mass estimates with
  luminosity \citep[e.g.,][]{shen10} or redshift, or a dependence on
  $\sigma_{BL}$ with luminosity or redshift. Both of these
  possibilities would decrease the dispersion in the scatter in the
  mass estimates when only probing a smaller range in $L$, or higher
  redshifts. In addition, \citet{marconi08} find a smaller scatter in
  the masses estimated using H$\beta$ when one corrects the
  reverberation mapped masses for radiation pressure. Because of the 
  possibility for a smaller scatter in the mass estimates, we perform our analysis with 
  both $\sigma_{BL}$ fixed to $\approx 0.2$ dex, and with
  $\sigma_{BL}$ as a free parameter.

  \subsection{The Posterior Distribution and Fitting Technique}

  \label{s-posterior}

  The technique for estimating the BHMF developed by KVF09 takes a 
  Bayesian approach for performing statistical inference, meaning that
  it calculates the probability distribution of the mass function,
  given the observed data. All the information regarding the BHMF and
  the parameters for the statistical model is contained within the
  posterior probability distribution, which is defined as the
  probability distribution of the model, given the observed
  data. KVF09 derived the
  posterior distribution for the statistical model described in the
  previous section. Denoting the model parameters for the shape of
  the BHMF as $\theta = (\pi, \mu, \Sigma, \gamma, \alpha_0, \alpha_m,
  \sigma_l)$, the posterior distribution is
  \begin{equation}
    p(\theta|{\rm FWHM},L,z) \propto p(\theta) \left[p(I=1|\theta) \right]^{-n} \prod_{i=1}^n 
    p({\rm FWHM}_i,L_{\lambda,i},z_i|\theta),
    \label{eq-thetapost}
  \end{equation}
  where the number of data points is $n$, $p(\theta)$ is the prior on
  $\theta$, and $p(I=1|\theta)$ is the probability as a function of
  $\theta$ that an BLQSO makes it into the SDSS DR3 catalogue. We note
  that if the dispersion in the mass estimates, $\sigma_{BL}$, is also
  treated as a free parameter, then $\theta$ also contains
  $\sigma_{BL}$. The joint
  distribution of ${\rm FWHM}, L,$ and $z$, $p(\log {\rm FWHM}_i, \log
  L_{\lambda,i}, \log z_i|\theta)$, is given by Equations (30)--(40)
  in KVF09, modified to use the mixture form for $p(L_{\lambda}|M_{BH})$. We use the the Mg II line at
  $1 < z < 1.6$, both the Mg II and the C IV line at $1.53 < z < 1.6$,
  and the C IV line at $z > 1.6$. We do not
  use H$\beta$ because we limit our analysis to $z > 1$. At $z
  \sim 0.8$ H$\beta$ is shifted into the water vapor bands, which
  tends to decrease the FWHM accuracy; Mg II is similarly affected
  at higher redshifts. In addition, we see systematic changes in the
  Mg II FWHM distribution above a redshift of 1.6, suggesting the
  presence of biases, which need further investigation (M. Vestergaard
  et al., in preparation). The
  posterior distribution for the BHMF normalization, given $\theta$
  and $n$, is given by Equation (16) in KVF09.

  The inclusion probability as a function of $\theta$ is calculated by
  averaging the SDSS DR3 selection function over the distribution of
  $L_{\lambda}$ and $z$ (see Eq.(46)--(49) in KVF09). In order
  to simplify our analysis we ignore the lower limit of ${\rm
    FWHM} > 1000\ {\rm km\ s^{-1}}$ on the emission line width for the
  SDSS DR3 sample. The distribution of ${\rm FWHM}$ for the SDSS DR3 falls
  off before reaching ${\rm FWHM} = 1000\ {\rm km\ s^{-1}}$, and it
  does not appear that imposing the lower limit results in a
  non-negligible fraction of the BLQSO population being
  missed. Therefore, any correction for the lower limit in ${\rm FWHM}$ is
  negligible, and we ignore it. In this case, the inclusion
  probability is
  \begin{eqnarray}
    p(I=1|\theta) & = & \frac{\Omega}{4\pi} \int_{L_{\lambda}=0}^{\infty} \int_{z=1}^{4.5}
    \frac{s(L_{\lambda},z)}{L_{\lambda} z \ln 10} \times \nonumber \\
    & & \sum_{k=1}^{K} \pi_k \left[ \sum_{j=1}^J \gamma_j N(\log
      L_{\lambda}|\bar{l}_{kj}(z),V_{l,kj})\right] N(\log
    z|\mu_{z,k},\sigma^2_{z,k})\ dz\ dL_{\lambda} \label{eq-incprob} \\
    \bar{l}_{kj}(z) & = & \alpha_{o,j} + \alpha_{m,j} \mu_{m,k} +
    \frac{\alpha_{m,j} \sigma_{mz,k}}{\sigma_{z,k}^2} \left( \log z -
      \mu_{z,k} \right) \label{eq-cmeanl} \\
    V_{l,kj} & = & \alpha_{m,j}^2 \sigma^2_{m,k} (1 -
    \rho_{mz,k}^2) + \sigma^2_{l,j}. \label{eq-cvarl}
  \end{eqnarray}
  Here, $\Omega = 1622\ {\rm deg^2}$ is the effective sky area of the
  SDSS DR3 sample \citep{dr3lumfunc}, $s(L_{\lambda},z)$ is the SDSS selection function,
  $N(x|\mu,\sigma^2)$ denotes a Normal distribution with mean $\mu$
  and variance $\sigma^2$, as a function of $x$, $\mu_{m,k}$ and
  $\mu_{z,k}$ are the mean values of $\log M_{BH}$ and $\log z$ for
  the $k^{\rm th}$ Gaussian function, respectively, and $\rho_{mz,k}$ is the
  correlation between $\log M_{BH}$ and $\log z$ for the $k^{\rm th}$
  Gaussian function. Note that $\bar{l}_k(z)$ and $V_{l,k}$ define the
  mean and variance in $\log L_{\lambda}$ at a given redshift for the
  $k^{\rm th}$ Gaussian function. The integral in Equation
  (\ref{eq-incprob}) is over $1 < z < 4.5$ because we have removed
  the sources at $z < 1$, and there are no useable broad emission
  lines at $z \gtrsim 4.5$.

  Equation (\ref{eq-incprob}) is in terms of the selection function
  with respect to the luminosity density. As mentioned above, we use
  the luminosity density at 1350\AA\ in this work. However,
  \citet{dr3lumfunc} report their selection function in terms of the
  $i$-band magnitude. We can convert the selection function of
  \citet{dr3lumfunc} to be in terms of the BLQSO power-law continuum
  $L_{1350}$ through the equation
  \begin{equation}
    s(L_{1350},z) = \int_{-\infty}^{\infty} s(i,z) p(i|L_{1350},z)\
    di, \label{eq-sfunc_convert}
  \end{equation}
  where $p(i|L_{1350},z)$ is the distribution of $i$-band magnitude at
  a given $L_{1350}$ and $z$, and $s(i,z)$ is the SDSS DR3 selection
  function in terms of $i$ and $z$. We model the distribution of $i$
  magnitudes at a given $L_{1350}$ in different redshift bins as a
  student's $t$ distribution with mean that depends linearly on $\log
  L_{1350}$:
  \begin{equation}
    p(i|L_{1350},z) = \frac{\Gamma[(\nu(z) + 1)/2]}{\Gamma(\nu(z)/2)
      \sigma_i(z) \sqrt{\nu(z) \pi}} \left( 1 + \frac{1}{\nu(z)}
    \left(\frac{i - A_i(z) - B_i(z) \log L_{1350}}
         {\sigma_i(z)}\right)^2\right)^{-(\nu(z) + 1) / 2}
    \label{eq-idist}
  \end{equation}
  The student's $t$ distribution converges to the normal distribution
  for $\nu \rightarrow \infty$; for finite $\nu$ the t-distribution is
  more heavy tailed than the normal distribution, and we have found it
  to better describe the distribution of $i$-magnitudes at a given
  $L_{1350}$ and $z$. The parameters $A_i(z), B_i(z), \sigma_i(z),$
  and $\nu(z)$ are fit by maximizing their posterior probability
  distribution, given the observed set of $i$ magnitudes and
  $L_{1350}$. The posterior distribution is given by
  inserting the assumed distributions into Equation (40) of
  \citet{kelly07a}, and for simplicity we assume a simple flux limit 
  of $i = 19.1$ for $z < 2.7$ and $i = 20.2$ for $z > 2.7$
  \citep[e.g.,][]{dr3lumfunc}. At $z < 2.7$ the redshift bins used in
  Equation (\ref{eq-idist}) have width $\Delta z = 0.1$, while at $z >
  2.7$ the redshift bins have width $\Delta z = 0.3$.

  In this work we constrain the parameters of our statistical model to be within certain
  limits, but in general assume a uniform prior on $\theta$. Our prior
  is uniform on the parameters within the limits $44.5 < \alpha_0 < 46.5,
  -1 < \alpha_m < 3, 0.1 < \sigma_l < 2, 7 < \mu_{m,k} < 10, \log 1
  < \mu_{z,k} < \log 4.5, 0.1 < \sigma_{m,k}, \sigma_{z,k} < 2,$ and
  $-0.98 < \rho_{mz,k} < 0.98$. We also assume a uniform prior on
  $\pi$ and $\gamma$, subject to the elements of $\pi$ and $\gamma$ summing to unity. The limits
  on $\alpha_m$ were chosen because we did not consider it realistic
  that $L_{1350}$ would depend on $M_{BH}$ outside of the range of
  dependencies spanning
  $L_{1350} \propto 1 / M_{BH}$ to $L_{1350} \propto M_{BH}^3$, and
  the limits on $\alpha_0$ were chosen to restrict the average value of
  $L / L_{Edd}$ to be within $0.01 < L / L_{Edd} < 1$ for BLQSOs
  with $M_{BH} = 10^9 M_{\odot}$, assuming a
  bolometeric correction of $C_{1350} = 4.3$.
  The limits on $\sigma_l$ were chosen because $\Delta \log L \approx 0.1$
  is comparable to the grid spacing on which the selection function
  was computed by \citet{dr3lumfunc}, and we did not consider it
  likely that the dispersion in $L_{1350}$ at a given $M_{BH}$ would
  be greater than $2$ dex. The limits on the BHMF parameters were chosen so as
  not to extrapolate the BHMF very far beyond the detectable range of
  $M_{BH}$. As such, in this work we limit our analysis of the BHMF to
  $M_{BH} \gtrsim 10^7 M_{\odot}$.

  As mentioned before, there exists the possibility that, in the range of $L$ and $z$ we probe, the
  uncertainty in the mass estimates may be smaller than the commonly
  quoted $\sim 0.4$ dex. If the
  error in the mass estimates is correlated with luminosity, then the
  variance in the mass estimates at a given luminosity and mass is
  reduced to
  \begin{equation}
    Var(\log \hat{M}_{BL}|L) = Var(\log \hat{M}_{BL}) (1 -
      \rho_{BL}^2).
    \label{eq-correrr}
  \end{equation}
  Here, $\hat{M}_{BL}$ denotes the broad line mass estimate, $Var(
  \log \hat{M}_{BL} )$ is the variance in the mass 
  estimates about the true mass, typically thought to be $\sim 0.4^2\
  {\rm dex}^2$, and $\rho_{BL}$ is the correlation with luminosity in
  the scatter in the mass estimates about the true mass. Because we probe
  a somewhat narrow range in $L_{\lambda}$, 
  when we estimate the parameter $\sigma_{BL}$ in Equation
  (\ref{eq-probvlm}), we are really estimating $2\sigma_{BL} \approx \sqrt{Var(\log
    \hat{M}_{BL}|L)}$, and we therefore need to also construct a
  prior for $\sigma_{BL}$. We do this by first noting that
  \citet{vest06} estimated the dispersion in the scatter in the broad
  line mass estimates about the reverberation mapping estimates using
  27 AGN. Therefore, the appropriate prior distribution for $Var(\log
  \hat{M}_{BL})$ is the posterior probability distribution of the
  variance in the mass estimates, given the reverberation mapping
  sample. Since we assume that the scatter in the mass estimates about
  the true mass is log-normal, the probability distribution of their
  variance follows using a standard result from Bayesian statistics,
  and is a scaled inverse $\chi^2$ distribution with 26 degrees
  of freedom \citep[e.g.,][]{gelman04}. However, there are currently
  no constraints on the value of $\rho_{BL}$, and we use a uniform
  prior on its value. Our prior on $\sigma_{BL}$ is then calculated by
  combining these two priors according to Equation
  (\ref{eq-correrr}). This results in a broad prior which peaks at
  $Var(\log \hat{M}_{BL}|L) \approx 0.4^2\ {\rm dex}^2$, and falls off slowly
  to zero as $Var(\log \hat{M}_{BL}|L) \rightarrow 0$ and $Var(\log
  \hat{M}_{BL}|L) \rightarrow 0.6^2$.

  In addition to the above constraints, we also impose the prior constraint that
  the number density of BLQSO SMBHs must never exceed the local number
  density of all SMBHs. In principle, this constraint could be
  violated if a large number of `wandering' black holes are
  present. These wandering black holes would be ejected from
  galactic nuclei in the late stages of a merger due
  to asymmetric emission of graviational radiation
  \citep[e.g.,][]{vol07}. However, this constraint would only be
  violated if the binary black hole system is ejected after or during the broad
  line quasar phase. While this would certainly be a very intriguing
  result, we do not test it here; indeed, our results are unaffected
  by this prior constraint as the estimated BHMF is always below the
  local value for all random draws from the posterior probability
  distribution.

  As described in KVF09, we do not work with Equation
  (\ref{eq-thetapost}) directly, but instead use a Markov Chain Monte Carlo (MCMC) sampler
  algorithm to obtain a set of random draws of $\theta$, distributed according to
  Equation (\ref{eq-thetapost}). Because the posterior distribution
  described by Equation (\ref{eq-thetapost}) is multimodal due to
  ambiguities in labeling the Gaussian functions, we label
  the BHMF Gaussian functions in order of increasing implied mean
  flux, and the $p(L|M_{BH})$ Gaussian functions in order of
  increasing mean luminosity at $M_{BH} = 10^9 M_{\odot}$. In
  addition, in order to make our MCMC sampling algorithm robust
  against additional possible multimodality, we include
  parallel tempering \citep[e.g.,][]{liu04} in our Markov Chain Monte Carlo
  (MCMC) algorithm to facilitate sampling from the different
  modes. For each value of $\theta$ we obtained from our MCMC random number
  generator, we also obtain
  a value of the BHMF normalization, $N$, by 
  drawing from a negative binomial distribution with parameters
  $n$ and $p(I=1|\theta)$ (KVF09). The random realizations of
  $\theta$ and $N$ then define a sample of random realizations of the BHMF obtained
  from the posterior probability distribution of the BHMF, given the observed
  set of mass estimates, luminosity densities, and redshifts. We
  ran our MCMC sampler for $2 \times 10^5$
  iterations each, keeping every $20^{\rm 
    th}$ iteration.

  \subsection{Robustness to Incorrect Assumptions Regarding the BLQSO
    BHMF and Eddington Ratio Distribution} 

  \label{s-incorrect_lcurve}

  Although we have assumed a flexible parameteric form for the BLQSO
BHMF and $p(L|M_{BH})$, this form is inconsistent with more
physically-motivated BLQSO $p(L|M_{BH})$, such as might be expected
from a power-law decay in the BLQSO accretion rate \citep[e.g.,][also
see discussion in \S~\ref{s-lcurves}]{yu05,stochacc}. Instead, our
assumed log-normal mixture form for $p(L|M_{BH})$ was motivated by
mathematical convenience, as some of the integrals necessary for
evaluation of the posterior distribution can be done analytically
(KVF09). If we were to use a form which required numerical
integration, we would have to perform over ten billion numerical integrals in
our MCMC sampler, which is computationally prohibitive. Moreover, we
also used the mixture form to allow flexibility 
in the estimated $p(L|M_{BH})$, so that our assumed form should be
able to approximate many different forms for $p(L|M_{BH})$. 

In order
to assess the impact of our chosen parameteric form on the inferred BHMF, we
simulated a data set where the distribution of Eddington ratios was
assumed to have a power-law form $p(\Gamma_{Edd}|M_{BH}) \propto
\Gamma_{Edd}^{-(1 + 1/\beta)}$ with $\beta = 2$ (see discussion in
\S~\ref{s-lcurves}). The distribution of bolometric corrections was
log-normal with geometric mean $C_{1350} = 5$ and dispersion of 0.2
dex, and the BHMF normalization was $N = 2 \times 10^6$; all other
aspects of the simulation were done in the same manner as described in
\S~6.1 of KVF09.

\begin{figure}
  \begin{center}
    \includegraphics[scale=0.33,angle=90]{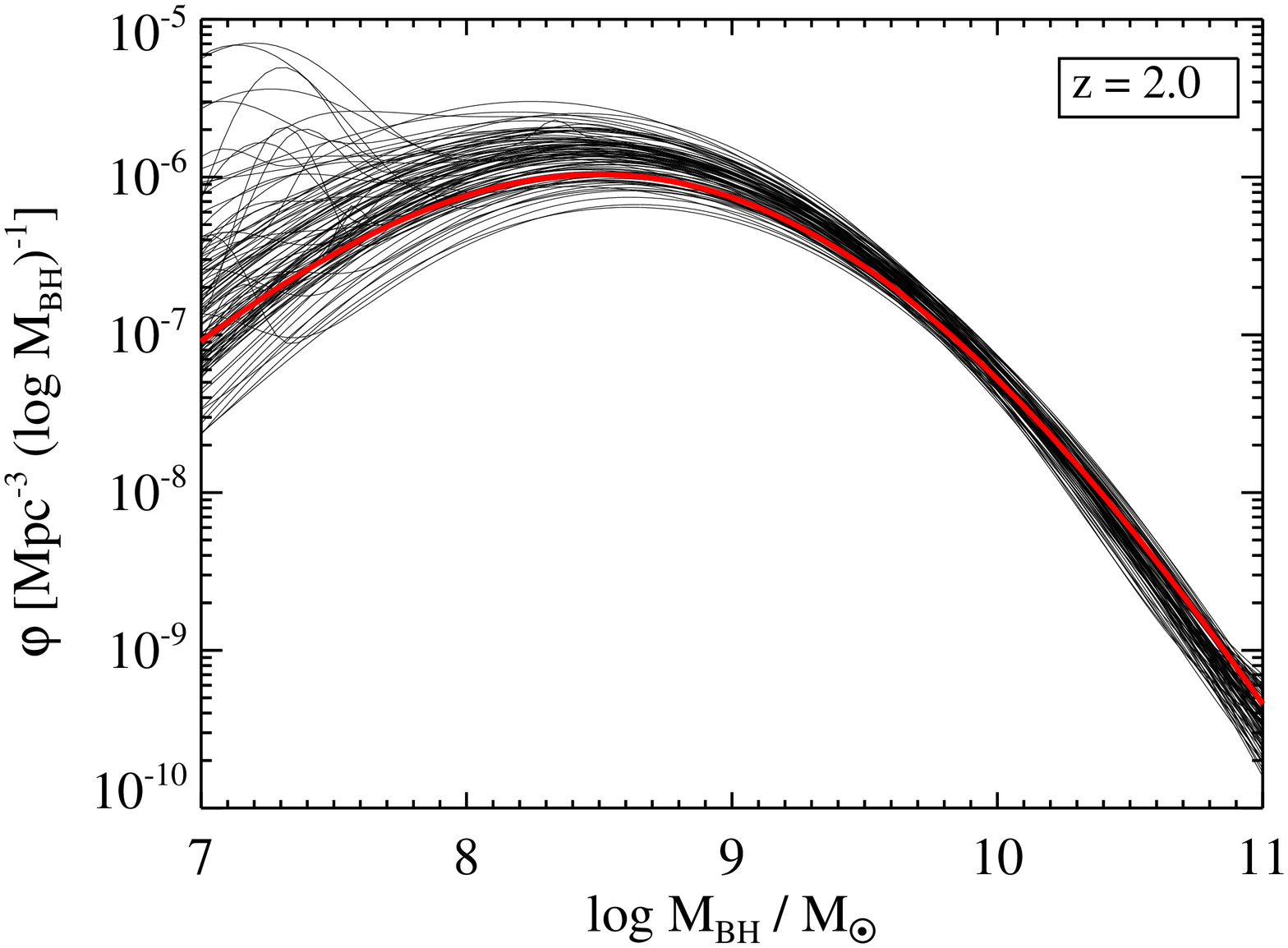}
    \includegraphics[scale=0.33,angle=90]{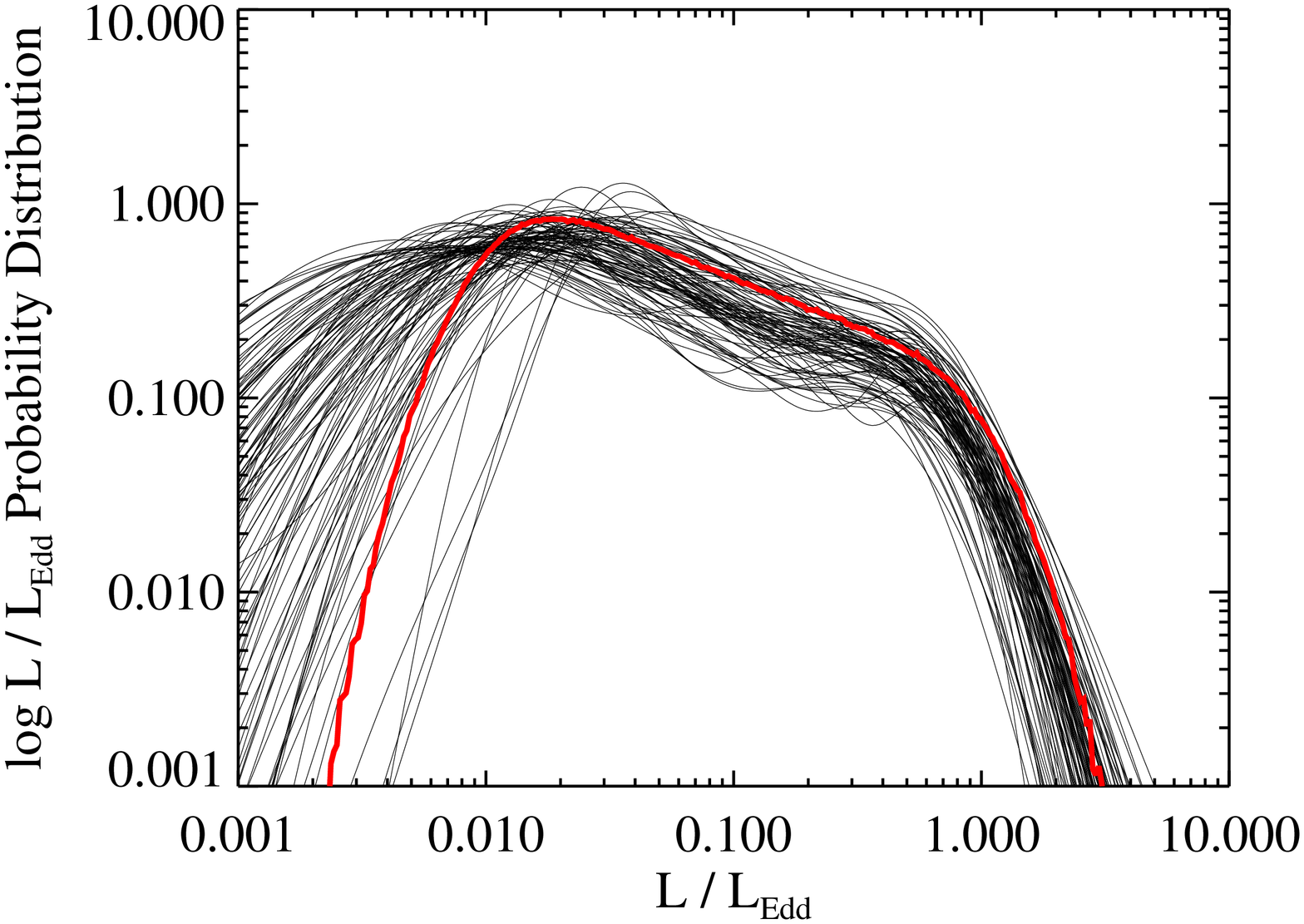}
    \caption{(a) True BHMF for the simulated sample described in
      \S~\ref{s-incorrect_lcurve}, having a power-law distribution of
      Eddington ratios over $0.01 < L / L_{Edd} < 1$ (thick red line),
      compared with the BHMF estimated 
      assuming a mixture of $J = 3$ log-normal distributions of Eddington
      ratios (thin black lines); each of the thin solid lines denotes a
      random draw from the probability distribution of the BHMF, given the
      observed data and assumptions outlined in \S~\ref{s-statmod2}. Here,
      and in all figures in this work, we plot 100 random draws of the
      function of interest, so the probability that the BHMF, say, has a
      certain value in a given range can be estimated by counting the number
      of random draws of the BHMF that fall within that range. For this
      example, the BHMF estimated assuming a mixture of log-normal
      distributions for $L / L_{Edd}$ is able to recover the true BHMF,
      implying that our mixture form is robust against mispecification of
      the parameteric model. (b) True Eddington ratio distribution for the
      same simulated sample (thick red line), compared with the estimated
      distribution assuming a mixture of $J = 3$ log-normal distributions
      (thin black lines). In this plot we have convolved the power-law form
      of the distribution of $L / L_{Edd}$ with the scatter in the
      bolometeric correction used in this simulation, to incorporate
      the error in the estimated Eddington ration distribution
      introduced from assuming a constant bolometeric 
      correction. The mixture of 
      log-normals form is able to adequately approximate
      the true distribution of $L / L_{Edd}$. \label{f-bhmf_sim}}
  \end{center}
\end{figure}

  The results are illustrated in Figure \ref{f-bhmf_sim}, where we
compare the true and estimated BHMF for the simulated sample at $z = 2$, and the
true Eddington ratio distribution with that inferred assuming the
statistical model described in \S~\ref{s-statmod2}. Here, and
elsewhere in this paper, in addition to a single `best-fit' mass
function, we also plot 100 random draws of the mass function from its
probability distribution, generated by our MCMC random number
generator. The spread and density of the random draws of the BHMF, and
any quantities derived from it, give a visual representation of the
uncertainty in these quantities, with the spread on the random draws
constraining the BHMF. Because we use 100 random draws, the
probability of a quantity falling within a certain area on a plot can
be estimated by counting the number of random draws that intersect
that area.  For this example, the BLQSO BHMF inferred assuming a
mixture of $J = 3$ log-normal distributions for $p(L_{\lambda}|M_{BH})$ is able
to recover the true BHMF and Eddington ratio distributions, at least
when the true distribution of $L / L_{Edd}$ is $p(\Gamma_{Edd})
\propto \Gamma_{Edd}^{-1.5}$. Although this test is far from
exhaustive, we consider it reasonable to conclude that our flexible
form for the BHMF and $p(L_{\lambda}|M_{BH})$ is able to accurately approximate
the true forms, and therefore our results are robust against errors
resulting from using an incorrect parameteric form for these
distributions.

  \section{RESULTS}

  \label{s-results}

  \subsection{Evaluating the Fit: How Uncertain are the Mass Estimates?}

  \label{s-postcheck}

\begin{figure}
  \begin{center}
    \includegraphics[scale=0.33,angle=90]{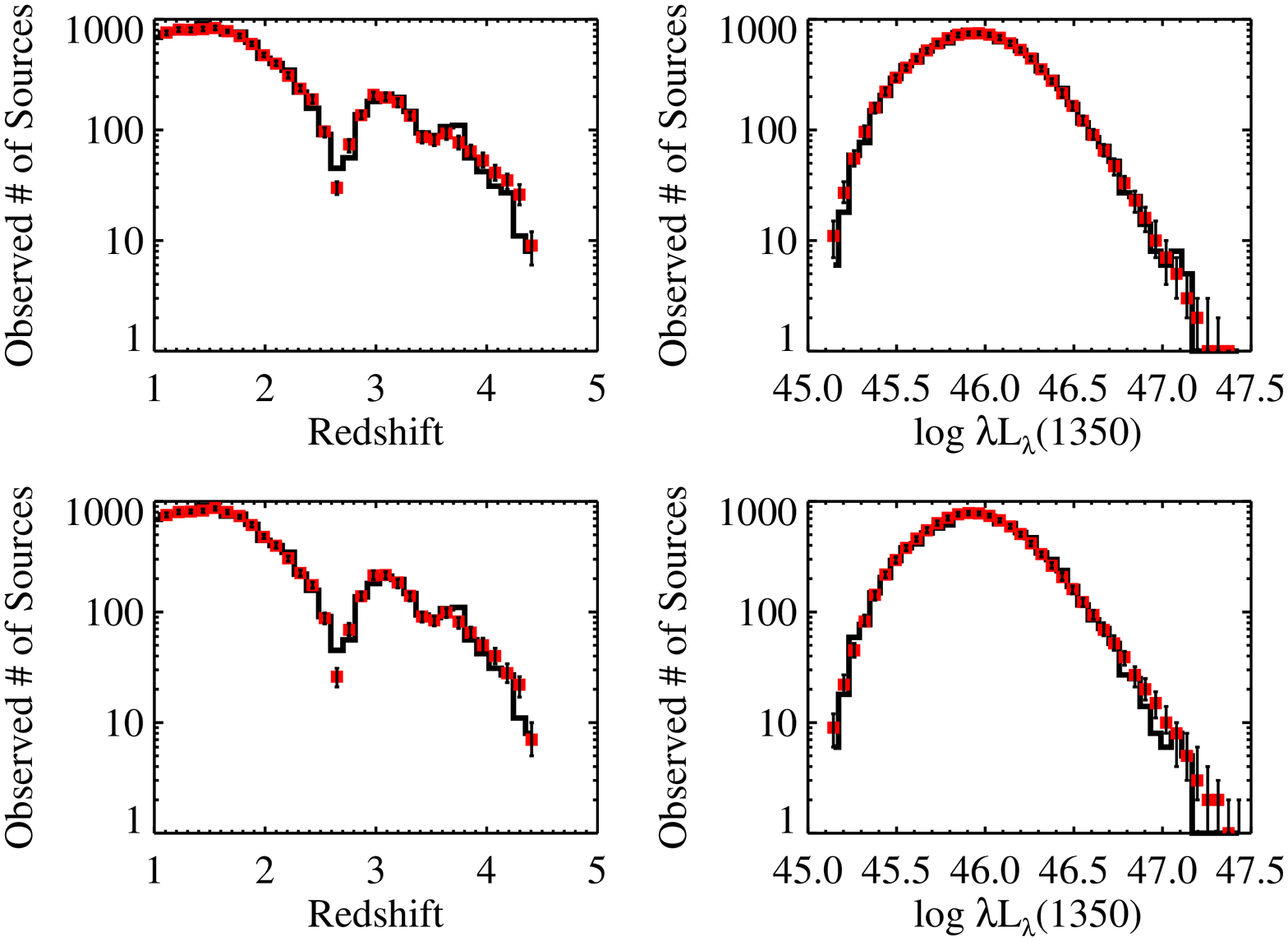}
    \includegraphics[scale=0.33,angle=90]{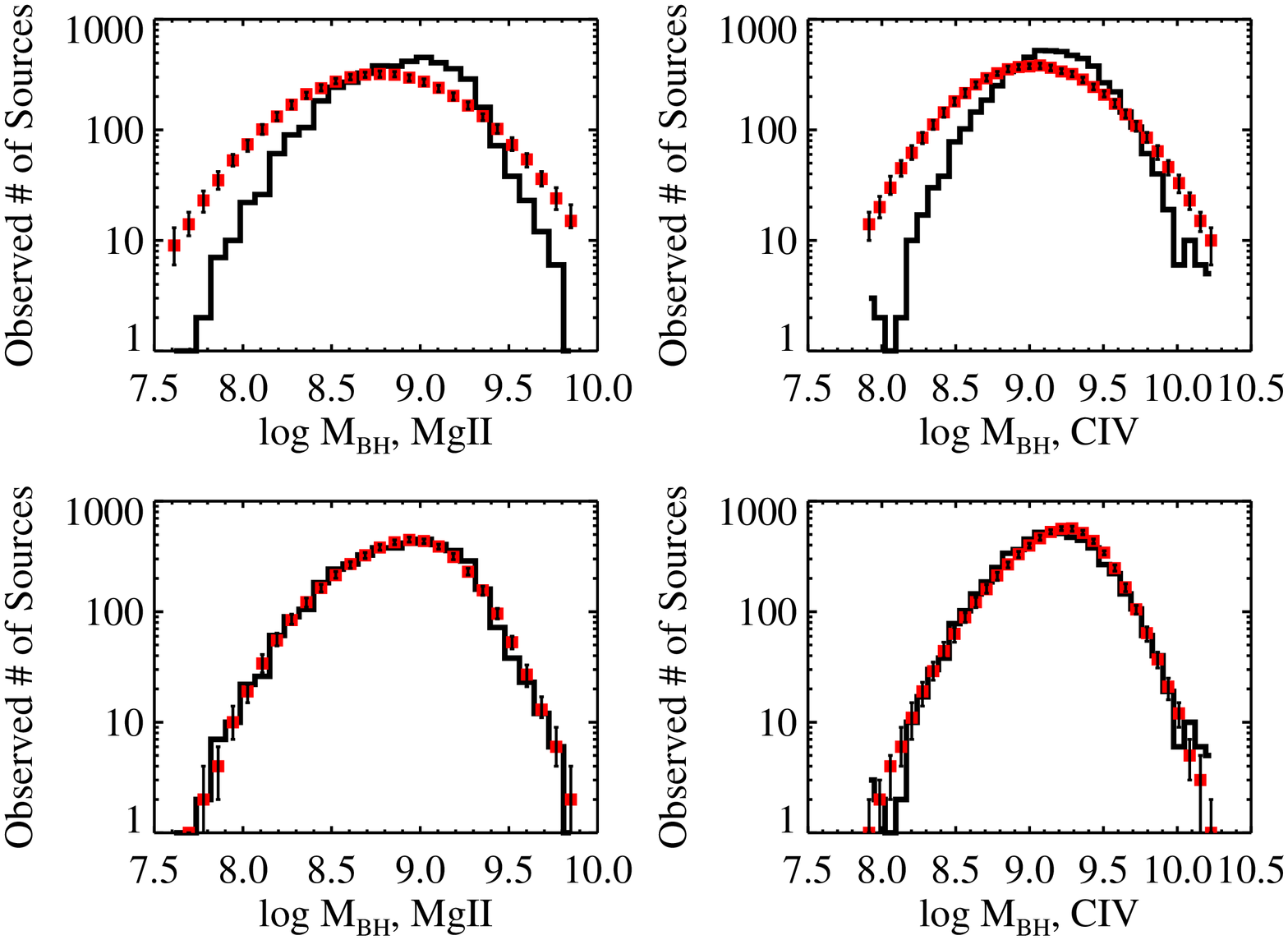}
    \caption{Distribution of $z, L_{1350},$ and the mass estimates for our
      sample (histograms), compared with the 
      distributions implied by our best-fit BHMF and distribution of
      $L_{1350}$ at a given $M_{BH}$ (red squares). In the top row of plots we fix the
      statistical error in the mass estimates to be $0.4$ dex, while 
      in the bottom row we allow it to be a free parameter. The error bars
      denote the $1\sigma$ uncertainties on the implied
      distributions, although often they are smaller than the red
      squares. The statistical model described in
      \S~\ref{s-statmod2} is able to reproduce the observed
      distributions only if we allow the standard deviation in the
      statistical error in the mass estimates to be a free parameter,
      implying that the standard value of $\sim 0.4$ dex
      is too large for this luminosity and redshift range. \label{f-postcheck1}}
  \end{center}
\end{figure}

  We used our Bayesian method to derive the BLQSO BHMF from the SDSS
  DR3 quasar sample, both holding the magnitude of the scatter in
  mass estimates fixed to $0.4$ dex, and treating the amplitude of the
  scatter as a free parameter. However, before discussing the results, we first
  evaluate how well the statistical model described in
  \S~\ref{s-statmod2} fits our sample. 
  In order to do this, we compare the distributions of redshift, luminosity, and broad line
  mass estimates of our sample to the
  distributions implied by our model, as described in
  KVF09. Figure \ref{f-postcheck1} compares the observed distributions of $z$,
  $\lambda L_{\lambda} ({\rm 1350\AA})$, and $\hat{M}_{BL}$ to those implied by the
  statistical model described in \S~\ref{s-statmod2}. The implied
  distributions were calculated by first simulating a sample of
  $M_{BH}$ and $z$ from a BHMF randomly output from the
  MCMC sampler. Then, for each value of $M_{BH}$, we simulated a value
  of $\lambda L_{\lambda} ({\rm 1350\AA})$ and mass estimate
  $\hat{M}_{BL}$. Finally, we 
  applied the selection function given by Equation
  (\ref{eq-sfunc_convert}) to the simulated data set. This was repeated
  for each realization of the BHMF obtained from the MCMC output, in order to
  account for the uncertainty in our estimated model parameters.

  We are not able to fit the mass estimate distributions if we keep
  the statistical scatter in the mass estimates fixed to 0.4 dex. In
  particular, this predicts a distribution of mass estimates that is
  too broad compared to the actual distribution. In fact, the observed
  dispersion in the mass estimates for our SDSS sample is $\approx
  0.35$ dex, smaller than that expected even if all objects had the
  same mass. However, if we allow
  the dispersion in the mass estimate error to be a free parameter, we
  are able to obtain a good match to the data. Our best-fit value for
  the scatter in the mass estimates at a given mass is $0.18 \pm 0.01$
  dex for Mg II and $0.13 \pm 0.01$ for C IV. Our result that the
  scatter in the mass estimates for high $L$ and $z$ must be smaller
  than $\sim 0.4$ dex is consistent with what has been found in
  previous work \citep{koll06,shen08,fine08}, and our best-fit values
  of the standard deviations in the mass estimate errors are consistent with
  the upper limits recently calculated by \citet{stein10b}.

  The origin of this smaller scatter is unclear, and there may be
  several different possibilities. One possibility, as mentioned
  earlier and by \citet{shen08} and \citet{shen10}, is that the error
  in the mass estimates may be correlated with luminosity. If this is
  true, then an error of $\sim 0.4$ dex represents the error in the
  mass estimates when averaging over a broad range in $L$, as was done
  for the reverberation mapping sample, while a
  smaller scatter of $\sim 0.15$ dex represents the error in the mass
  estimates when one is limited to a more narrow range in $L$, as the
  SDSS quasar sample is. It is unclear why the error in the mass
  estimates would be correlated with luminosity, but one possible
  source unaccounted for is
  radiation pressure. \citet{marconi08} argue that virial
  mass estimates should be corrected for radiation pressure. They find
  a correction that implies a steeper dependence on luminosity than
  $\hat{M}_{BL} \propto L^{0.5}$, especially for sources with high $L
  / L_{Edd}$. Under their model, if one does not correct for radiation
  pressure then one will tend to underestimate the mass with
  increasing $L$, producing a correlation between the error in the
  mass estimates with luminosity, and possibly producing a smaller
  scatter in the mass estimate errors over a narrow range in
  luminosity. It is interesting to note that 
  \citet{marconi08} find a smaller scatter in the mass estimates of $\sim
  0.2$ dex when correcting for radiation pressure for the
  reverberation mapped sample, which covers a larger range in
  luminosity. However, more work is need to 
  understand the importance of radiation pressure, and if it can
  produce the observed smaller scatter in the mass estimates over the
  range in luminosity we probe.

  Another possibility is that the error in the mass estimates may not
  be correlated with $L$, but the dispersion in the errors may
  decrease with increasing $L$ or $z$. Unfortunately, the number of
  AGN with $M_{BH}$ estimated from reverberation mapping is too small
  to test this, and dominated by sources at lower $L$ and $z$. The third
  possibility is that the virial mass estimates are biased at high $L$
  and $z$, at least for Mg II and C IV, and are only marginally related to the actual
  masses. The $R$--$L$ relationship is well established for H$\beta$,
  and does not require a large extrapolation to the luminosities
  probed in our sample, and thus we do not expect H$\beta$-based mass
  estimates to be significantly biased in this range \citep{vest07}. The situation
  is less clear for Mg II and C IV. There is only one reliable time
  lag for the Mg II emission line \citep{metz06}. An $R$--$L$
  relationship has been estimated for the C IV line over a broad range
  in luminosity and redshift \citep{kaspi07}, including those probed
  in our study, and the $R$--$L$ relationship is similar for both
  H$\beta$ and C IV. Unfortunately, the C IV $R$--$L$ relationship is
  estimated from only eight data points, and further work is needed in order to
  understand the Mg II- and C IV-based mass estimates and their errors.

  Throughout the rest of this work will we focus on the results
  obtained from allowing the dispersion in the mass estimate error to
  be a free parameter, a value of $\sim 0.4$ dex is clearly ruled
  out. However, we note that we have performed the same analysis for
  both cases, and while the quantitative details change, 
  the scientific conclusions are unaffected by treating the amplitude
  of the scatter as a free parameter. The only exception is that we
  infer a much more narrow distribution of Eddington ratio if we fix
  the amplitude of the scatter in the mass estimates to be $\sim 0.4$
  dex. In addition, the results 
  reported in this section highlight the need for more reverberation
  mapping studies in order to better understand the nature of the
  errors in the mass estimates.

  \subsection{The Black Hole Mass Function for Broad Line AGN}

  \label{s-bhmf}

\begin{figure}
  \begin{center}
    \scalebox{0.7}{\rotatebox{90}{\plotone{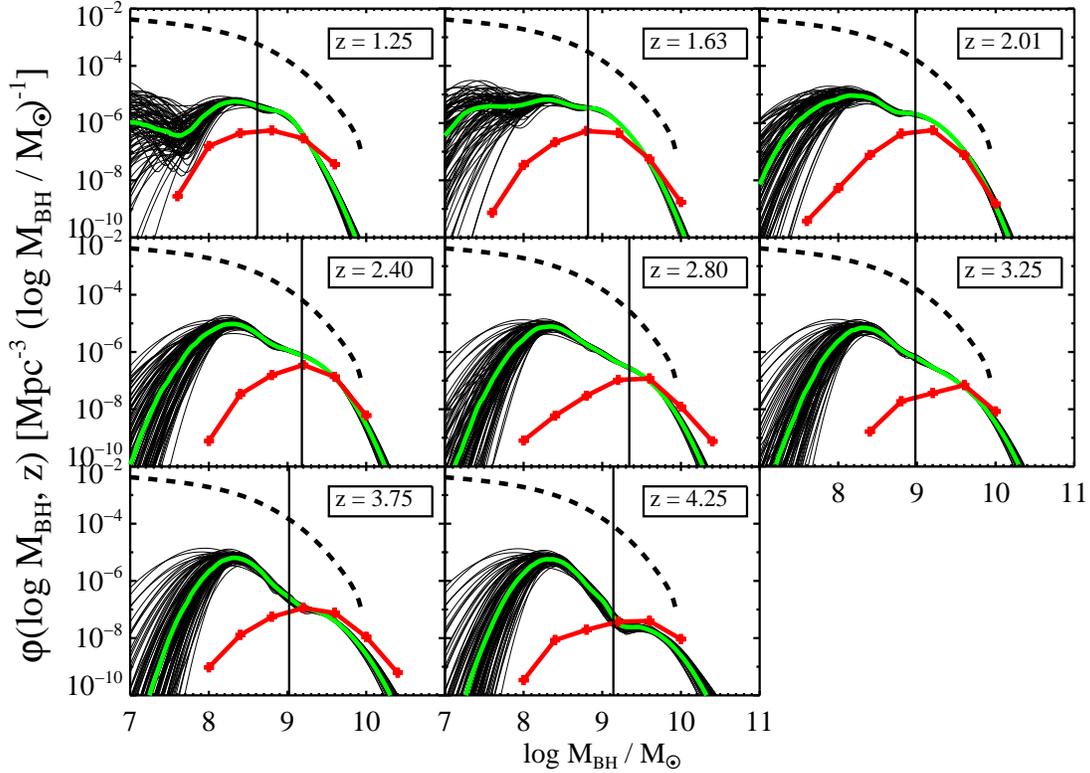}}}
    \caption{BLQSO BHMF (thin solid lines) obtained using our Bayesian
      approach, compared with the local BHMF fo all SMBHs (dashed
      line), and the BHMF from \citet[][solid red line with
      points]{vest08}; as in Figure \ref{f-bhmf_sim}, each thin solid
      line denotes a random draw of the BHMF from its probability
      distribution. The thick green line is the median of the 
      BHMF random draws, and may be considered our `best-fit'
      estimate. The vertical line marks the mass at which the SDSS DR3
      sample becomes $10\%$ complete. \label{f-bhmf}} 
  \end{center}
\end{figure}

  Figure \ref{f-bhmf} shows the BLQSO BHMF at
  several redshifts. Our estimated BHMF is compared
  with an estimate of the local mass function of all SMBHs, and the
  BHMF reported by \citet{vest08}, obtained from binning up the broad
  line mass estimates. Following
  \citet{merloni08}, the local BHMF was 
  computed to be near the middle of the uncertainty range reported by
  \citet{shank09} by convolving a Schechter function with a
  log-normal distribution with standard deviation 0.3 dex, chosen to
  be consistent with the scatter about the $M_{BH}$--$\sigma$ relationship; the
  parameters for the Schechter function are those reported by
  \citet{merloni08}. Also, we note that early type
  galaxies dominate the local BHMF at $ M_{BH} \gtrsim 4 \times 10^7
  M_{\odot}$ \citep{yu08}. 

\begin{figure}
  \begin{center}
    \includegraphics[scale=0.33,angle=90]{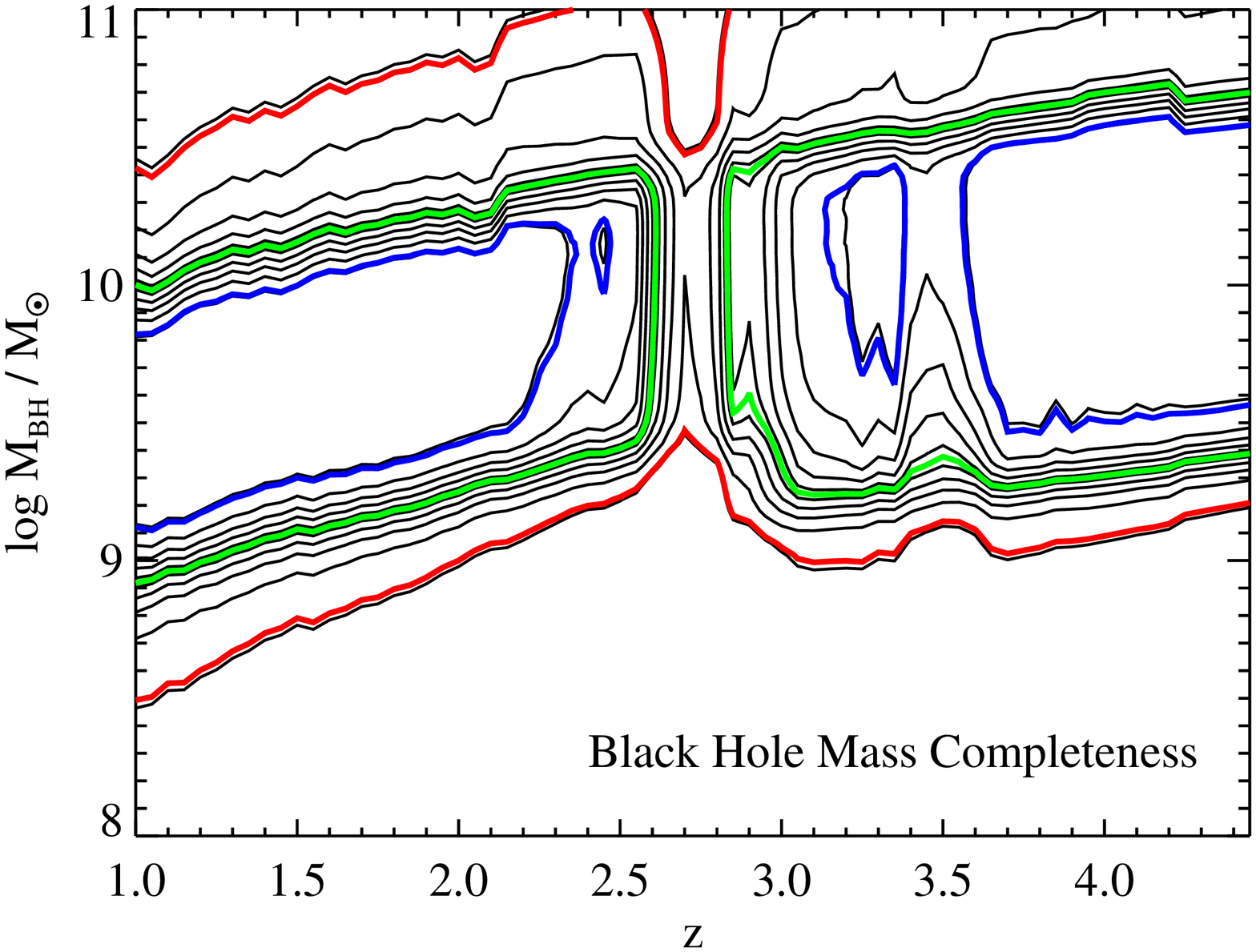}
    \includegraphics[scale=0.33,angle=90]{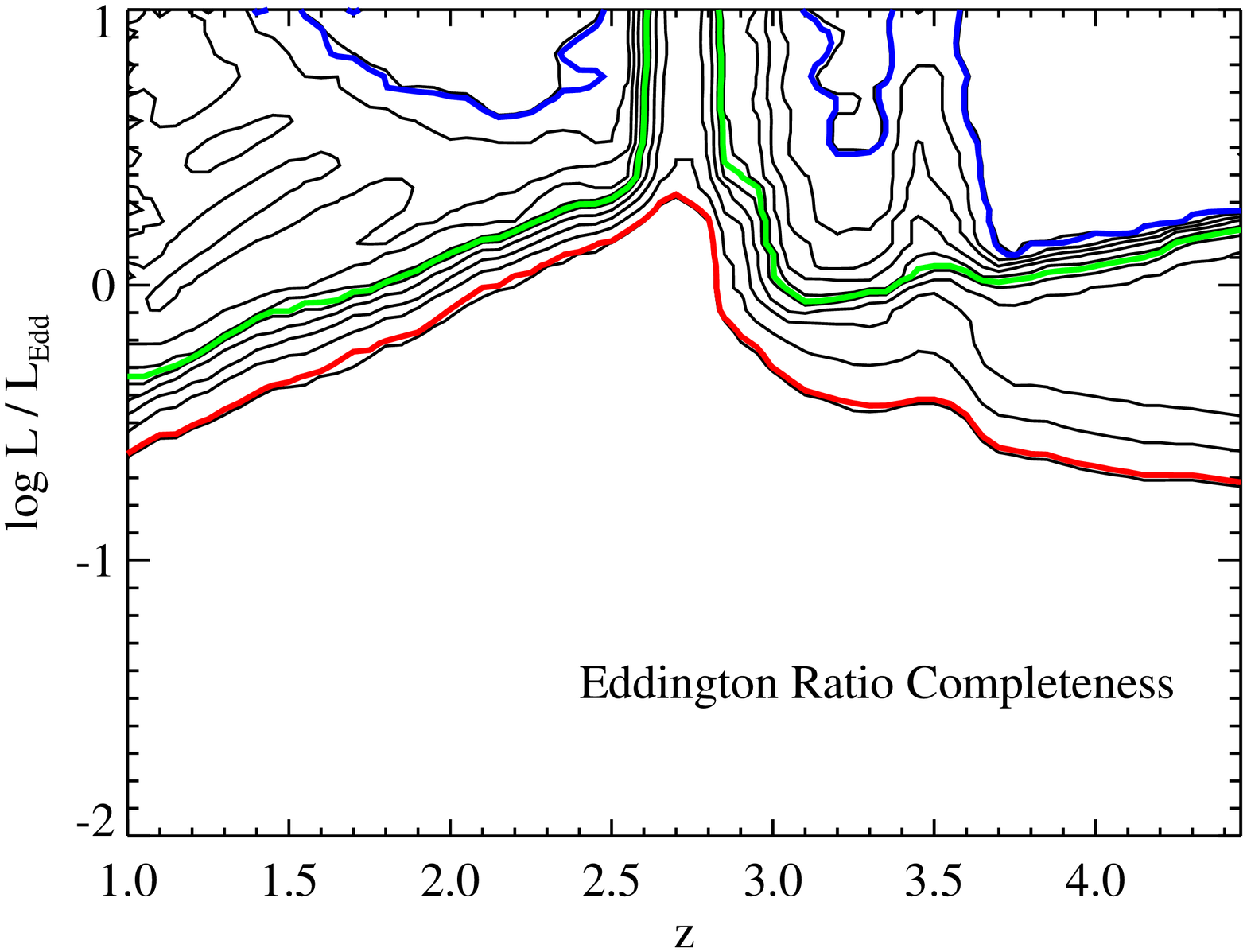}
    \caption{Estimated completeness in black hole mass (left) and
      Eddington ratio (right) for the SDSS
      DR3 quasar sample of \citet{dr3lumfunc}, calculated using our
      assumed distribution of luminosity at a given black hole mass
      (Eq.[\ref{eq-problm}]). The red, green, and blue lines denote 
      the $10\%, 50\%,$ and $90\%$ completeness levels, respectively. The SDSS
      sample is highly incomplete at $M_{BH} \lesssim 10^9
      M_{\odot}$ and $L / L_{Edd} \lesssim 0.5$. Note that the
      incompleteness at the highest masses is due to the upper flux
      limit of the SDSS. \label{f-completeness}}
  \end{center}
\end{figure}

  In order to focus on the region of the BHMF that is robust against uncertainties in
  the selection function, as well as against uncertainty on the
  Eddington ratio distribution, we estimate the black hole mass
  completeness for our sample as a function of $z$. Our best estimate of the SDSS
  completeness as a function of $M_{BH}$ and $z$ is shown in Figure
  \ref{f-completeness}, and the $10\%$ completeness limit is marked by
  a vertical line in Figure \ref{f-bhmf}. The black hole mass completeness depends on
  both the completeness in luminosity, and the assumed distribution of
  $L$ at a given $M_{BH}$. As can be seen from Figure
  \ref{f-completeness}, at $z \gtrsim 2$ the SDSS quasar sample is only
  $\approx 10\%$ complete at $M_{BH} \sim 10^9 M_{\odot}$, becoming
  more incomplete at lower masses. At masses much lower than $M_{BH}
  \sim 10^9 M_{\odot}$, the estimated mass function almost completely
  depends on extrapolation from the set of BHMFs and Eddington ratio
  distributions that fit the observed data well, constrained by our
  assumed parameteric forms. Therefore, we stress that below $M_{BH}
  \sim 10^9 M_{\odot}$ the BLQSO BHMF must be interpreted with
  caution, and in this work we will try to focus on what we can infer
  from the high mass end of the mass function.

  We also estimate the completeness of our sample as a function of $L
  / L_{Edd}$, assuming a constant bolometric correction of $C_{1350} =
  4.3$. The estimated completeness depends on the selection function
  and the estimated BHMF, as the selection function depends on
  luminosity, which is defined by the BHMF at a given $L /
  L_{Edd}$. The Eddington ratio completeness is also shown in Figure
  \ref{f-completeness}. The SDSS is only $\sim 50\%$ complete for
  sources radiating at the Eddington limit, and $\lesssim 10\%$ complete for
  sources radiating at $\lesssim 10\%$ of Eddington. This heavy incompleteness
  is due to the fact that, for our estimated BLQSO BHMF, half of
  BLQSOs have black holes that are not massive enough to make the SDSS
  flux limit even if they radiate at Eddington. We further discuss the
  Eddington ratio distribution in Section \S~\ref{s-eddrat}.

  The difference between the binned estimate of the BLQSO BHMF,
  calculated from the estimate of \citet{vest08}, and our estimate, is the
  result of the difference in statistical methodology employed by
  \citet{vest08} and our work. First, the statistical error in
  the broad line mass estimates results in a broader inferred BHMF
  when one simply bins up these estimates
  \citep[e.g.,][KVF09]{kelly07b}. Second, the $1 / V_a$ technique
  corrects for incompleteness in flux, but not black hole mass. As a
  result, the $1 / V_a$ corrections only partially correct for
  incompleteness in $M_{BH}$, and the estimated binned BHMF will still
  suffer from incompleteness, especially at the low mass end. The two
  effects combined result in a systematic shift in the estimated BHMF
  toward higher $M_{BH}$ \citep[][KVF09]{shen08}. However, the
  Bayesian approach outlined in KVF09 is able to self-consistently
  correct for these two effects in a statistically rigorous manner,
  conditional on the survey selection function and assumptions
  implicit within the statistical model. In spite of these differences
  in methodology, our estimated BHMF agrees fairly well with that of \citet{vest08}
  at the high mass end, considering the differences in the
  methodology, and the two do not strongly diverge until values of
  $M_{BH}$ where the SDSS is highly incomplete. The poorer agreement
  between the two estimates 
  at $z < 2$ is due to the fact that the BHMF is estimated using the
  Mg II emission line in this redshift range, which we find to have a
  higher statistical error than that of C IV. As a result, the
  our correction to the BHMF due to the error in the mass estimates is
  greater at these redshifts.

\begin{figure}
  \begin{center}
    \scalebox{0.7}{\rotatebox{90}{\plotone{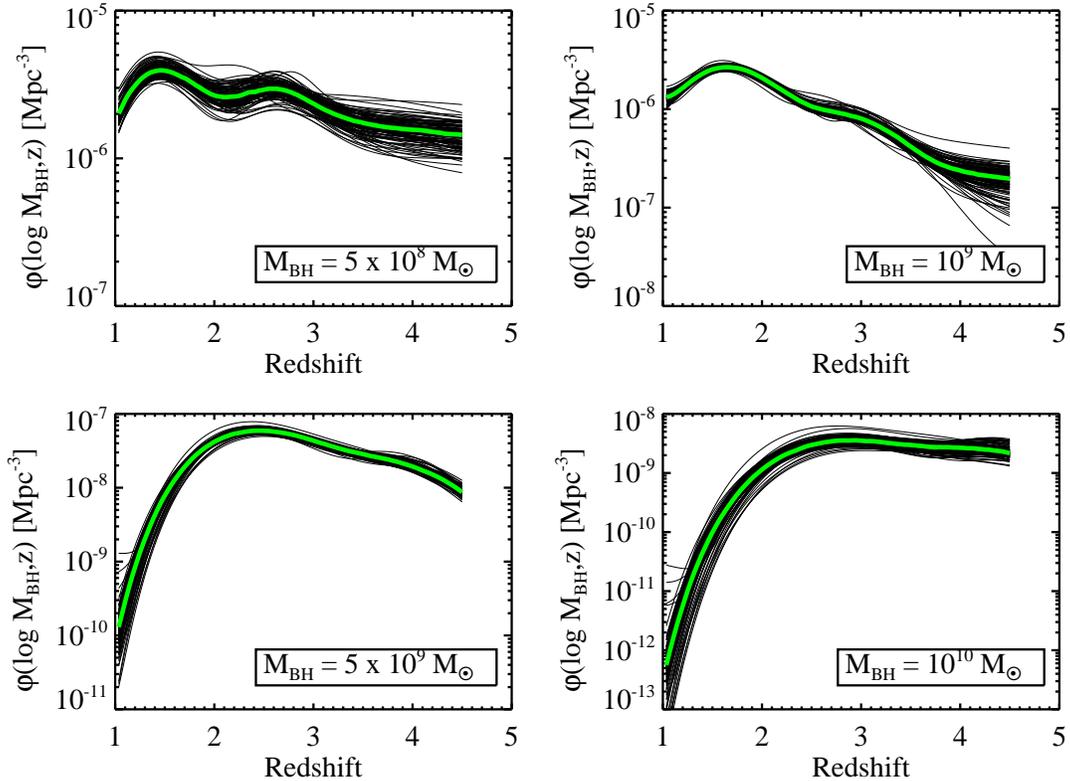}}}
    \caption{Evolution in the number density of BLQSO SMBHs for four
      different values of $M_{BH}$. A larger 
      fraction of the most massive SMBHs are seen as BLQSOs at higher
      redshift, an effect commonly referred to as black hole
      downsizing. Downsizing of BLQSO SMBHs has also been seen in the
      LBQS \citep{vest09}. Symbols are as in Figure \ref{f-bhmf}. The
      wiggles seen in the lowest mass bin are an artifact of fitting a
      Gaussian mixture model, unlikely to be real, and probably due to small
      unaccounted for errors in the selection function, which can
      become magnified due to the the fact that we are 
      incomplete in this mass bin. \label{f-downsize}}
  \end{center}
\end{figure}

  In Figure \ref{f-downsize} we show the evolution in the comoving
  number density of SMBHs in BLQSO at four different masses. As is
  evident, the number density 
  of higher mass BLQSO SMBHs peaks at higher redshift, where the
  number density of BLQSO SMBHs of $M_{BH} \sim 5 \times 10^8 M_{\odot}$
  peaks at $z \sim 1.5$, and the number density for $M_{BH} \sim 5
  \times 10^{9} M_{\odot}$ peaks at $z \sim 2.5$. This result is
  consistent with what has commonly been referred to in the literature
  as black hole `downsizing', where the most massive SMBHs are active,
  and grow, at earlier epochs than lower mass black holes, an effect
  also seen by \citet{vest09} and \citet{stein10a}.

  \subsection{The Broad Line Quasar Black Hole Mass Density, and
    Constraints on their Duty Cycle and Average Lifetime}

  \label{s-lifetime}

\begin{figure}
  \begin{center}
    \scalebox{0.7}{\rotatebox{90}{\plotone{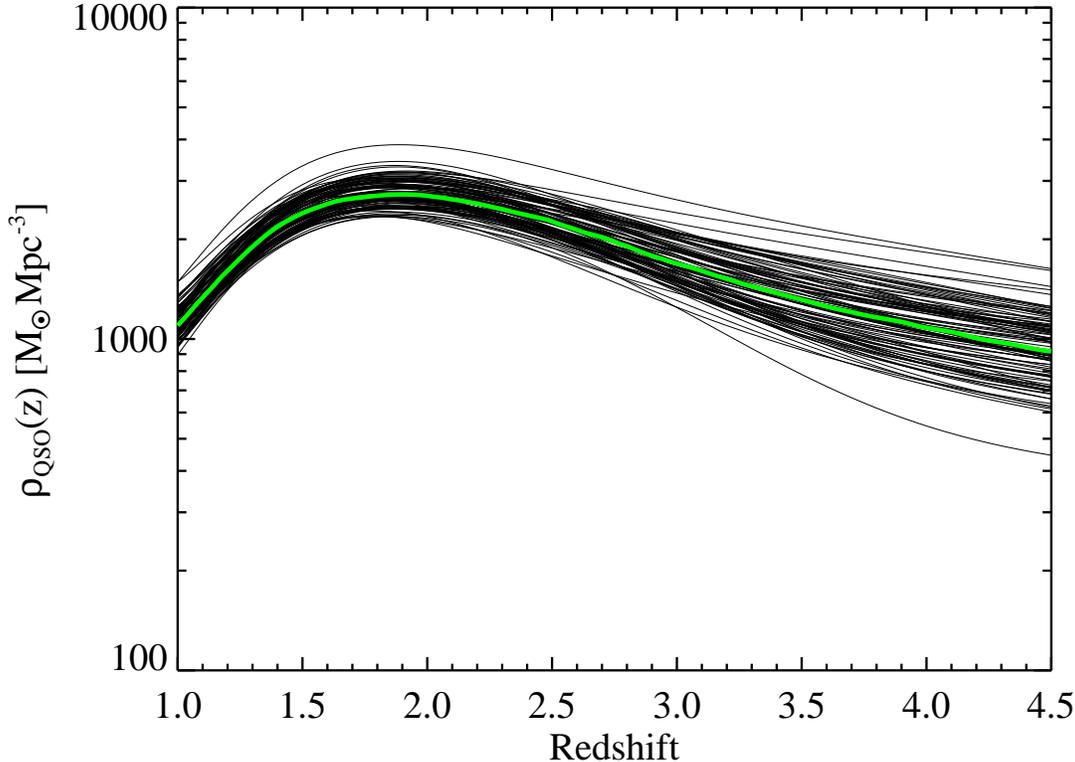}}}
    \caption{Evolution in the mass density of SMBHs seen as
      BLQSOs, symbols as in Figure \ref{f-bhmf}. For comparison, recent estimates of the local mass
      density of all SMBHs suggest $\rho_{BH}(z=0) \sim 4 \times 10^5
      M_{\odot} {\rm Mpc^{-3}}$ \citep[e.g.][]{yu08}. The mass density
      of SMBHs in BLQSOs peaks at $z \sim 2$. \label{f-massdensity}}
  \end{center}
\end{figure}

  In Figure \ref{f-massdensity} we show the comoving mass density of
  SMBHs that reside in BLQSOs, as a function of redshift,
  $\rho_{QSO}(z)$. The peak in the cosmic mass
  density of BLQSO SMBHs occurs at $z \sim 2$. Our constraint on the location of the peak in
  $\rho_{QSO}(z)$ is consistent with the peak in the mass density
  derived from the Large Bright Quasar Survey and high-$z$ sample of
  \citet{fan01a}, as calculated by \citet{vest09}. Previous work has not found any
  evidence for evolution in the Eddington ratio distribution at $z >
  1$ for the most massive BLQSOs
  \citep[e.g.,][]{mclure04,vest04,koll06,vest09}. If there is no
  significant evolution in the Eddington ratio distribution at $z > 1$
  for these systems, then we would expect the peak in their luminosity
  density to coincide with the peak in their black hole mass
  density. The luminosity density of quasars peaks at $z \sim 2$
  \citep[e.g.,][]{wolf03,hopkins07}, matching the peak in black hole mass density observed in our work.

\begin{figure}
  \begin{center}
    \scalebox{0.7}{\rotatebox{90}{\plotone{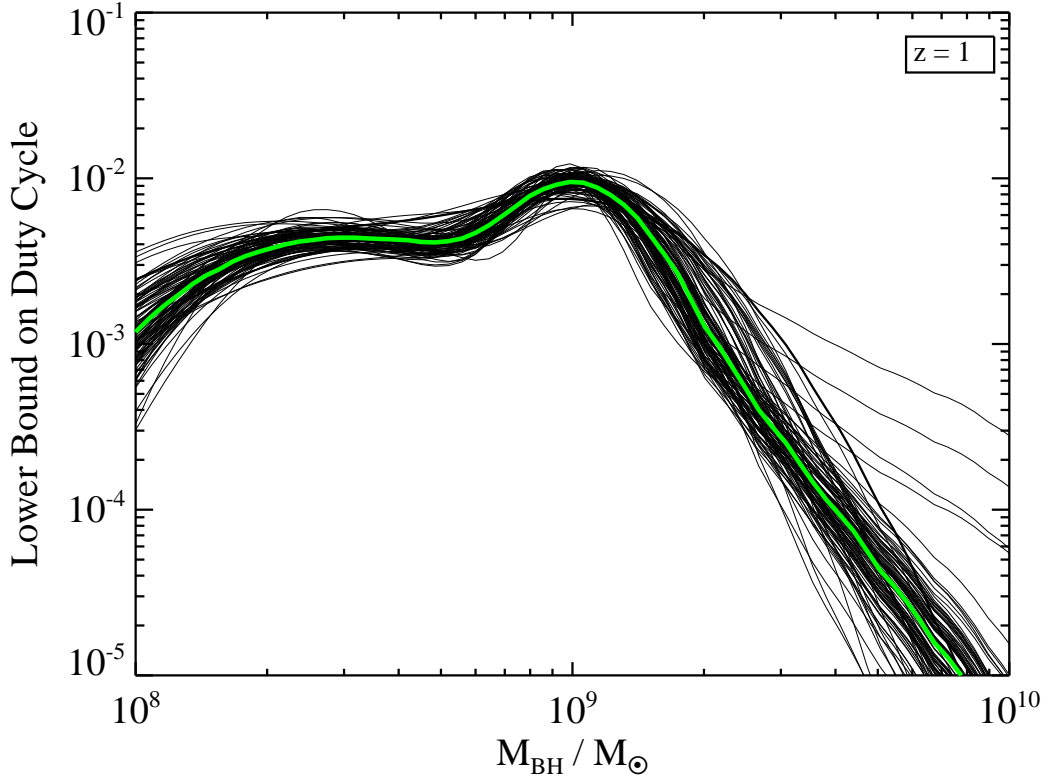}}}
    \caption{Lower bound on the duty cycle for BLQSO activity at $z = 1$ as a function
      of $M_{BH}$, symbols as in Figure \ref{f-bhmf}. The duty cycle for $M_{BH} \sim 10^9
      M_{\odot}$ BLQSO SMBHs at $z \sim 1$ is $\delta \gtrsim 0.01$,
      falling to $\delta \gtrsim 10^{-5}$ for $M_{BH} \sim 10^{10}
      M_{\odot}$. The decrease in the duty cycle with increasing
      black hole mass is likely due to either cosmic downsizing or a
      shorter BLQSO phase for the most massive SMBHs. \label{f-dutycycle}}
  \end{center}
\end{figure}

  We can use the quasar BHMF to place constraints on the fraction of
  black holes that are seen in the BLQSO phase as a function of
  $M_{BH}$; this quantity is commonly referred to as the BLQSO `duty
  cycle'. Denote the duty cycle as $\delta(M_{BH},z)$. Then
  \begin{equation}
    \delta(M_{BH},z) \equiv
    \frac{\phi_{QSO}(M_{BH},z)}{\phi_{BH}(M_{BH},z)}. \label{eq-dutycycle}
  \end{equation}
  Here, $\phi_{BH}(M_{BH},z)$ is the BHMF of all SMBHs at a given
  redshift. Ignoring mergers of SMBHs, we can compute a lower-limit to
  the duty cycle by comparing the BHMF at a certain redshift with the
  local SMBH number density, since $\phi_{BH}(M_{BH},z) \leq
  \phi_{BH}(M_{BH},0)$. In Figure \ref{f-dutycycle} we compute the
  lower limit of the BLQSO duty cycle at $z = 1$ for SMBHs with
  $M_{BH} > 5 \times 10^8 M_{\odot}$. The duty cycle at $z = 1$ is
  constrained to be $\delta \gtrsim 0.01$ at $M_{BH} \sim 10^{9}
  M_{\odot}$, falling steeply to $\delta \gtrsim 10^{-5}$ at $M_{BH} \sim
  10^{10} M_{\odot}$. The decrease in the duty cycle with increasing
  black hole mass may be another reflection of SMBH downsizing, with
  the most massive BLQSO SMBHs being active at earlier cosmic
  epochs. Alternatively, it may be due to the fact that the most
  massive SMBHs spend a shorter amount of time in the broad line
  phase, as expected from some simulations of black hole feedback
  \citep[e.g.,][]{faintend}. However, we note that the local number
  density of the most massive SMBHs is poorly constrained and subject
  to considerable systematic uncertainty 
  \citep{lauer07}, and therefor duty cycle of BLQSOs with $M_{BH} \sim
  10^{10} M_{\odot}$ may be subject to considerable systematic
  error. Indeed, the observed fall-off in the lower limit on the duty cycle for the most
  massive systems may simply be due to the fact that we are
  overestimating the number density of local SMBHs.

  The quasar lifetime is not well constrained empirically
  \citep[e.g.,][]{martini04}, but we can use our estimated BLQSO BHMF
  to estimate the BLQSO lifetime. For simplicity, we assume that all
  BLQSOs of a given mass have a single lifetime,
  $t_{BL}$. Furthermore, if a SMBH undergoes multiple episodes of BLQSO activity,
  then we assume that $t_{BL}$ is the same for each episode. These assumptions are
  unlikely to be true, but for the purposes of our work we may still
  think of $t_{BL}$ as a `typical' BLQSO lifetime. Because the duty
  cycle is the probability of observing a SMBH as a BLQSO at a given
  $M_{BH}$ and redshift we can relate $t_{BL}$ to
  $\delta(M_{BH},z)$. The probability of observing a SMBH as a BLQSO
  at redshift $z$ is the product of the probability that a SMBH
  becomes a BLQSO along our line of sight at some point in its evolution with the probability
  that that SMBH is a BLQSO between cosmic ages $t(z) - t_{BL}$ and
  $t(z)$, normalized by the probability that that SMBH is a BLQSO at
  $t \leq t(z)$. The normalization results from the additional
  assumption that if a SMBH of mass $M_{BH}$ exists at $t(z)$, and if
  it goes through at least one BLQSO episode as some point in its growth, then it had to have
  gone through at least one BLQSO episode at $t < t(z)$ in order to grow to $M_{BH}$ by
  $t(z)$. This is a reasonable assumption, at least for the most
  massive system, since all SMBHs at $t(z)$ had to grow from much
  lower mass seeds, and if BLQSO activity occurs for a SMBH at some
  point in its life, BLQSO activity would have occurred during this
  growth period. In other words, the assumption implicit in the
  normalization correction is that SMBH seeds of mass, say, $M_{BH}
  \sim 10^9 M_{\odot}$, do not simply emerge and then undergo BLQSO
  activity at a cosmic epoch later than $t(z)$. The practical result
  of this assumption is that the BLQSO duty cycle of, say, SMBHs with
  $M_{BH} \sim 10^9 M_{\odot}$, can decrease from $\delta \sim 1$ at
  $z \sim 7$ to $\delta \sim 10^{-3}$ at $z \sim 1$.

  Denote the time that a BLQSO phase is initiated in a SMBH as
  $\tau$. Then, the probability that a SMBH BLQSO is seen between
  $t(z) - t_{BL}$ and $t(z)$ is the same as the probability that the
  time that a BLQSO phase was initiated for that source occured at
  $t(z) - t_{BL} < \tau < t(z)$. Note that this allows for multiple
  BLQSO episodes, as multiple phases of BLQSO episodes simply alter
  the probability of observing a SMBH as a BLQSO at a
  certain redshift. The duty cycle is thus related to $t_{BL}$ as
  \begin{equation}
    \delta(M_{BH},z) = Pr(BLQSO|M_{BH}) \left[ \frac{ \int_{t(z) - t_{BL}}^{t(z)}
    p(\tau|M_{BH},BLQSO)\ d\tau }{ \int_{0}^{t(z)}
    p(\tau|M_{BH},BLQSO)\ d\tau } \right],
    \label{eq-dcycle}
  \end{equation}
  where, $Pr(BLQSO|M_{BH})$ is the probability that a SMBH with a mass
  of $M_{BH}$ is a BLQSO at some
  point in its evolution, and $p(\tau|M_{BH},BLQSO)$ is the probability
  distribution of BLQSO episode initiation time for a given black hole
  mass. Note that $Pr(BLQSO|M_{BH})$ is not the probability that an
  object is currently seen as a BLQSO, but is the probability that it
  appears as a BLQSO to an observer on Earth at some point in its
  life. As mentioned above, $p(\tau|M_{BH},BLQSO)$ is sufficiently 
  general to include both multiple and single episodes of BLQSO
  activity, although the actual form of $p(\tau|M_{BH},BLQSO)$ will
  depend on the distribution of the number of BLQSO episodes a SMBH goes
  through. All values of $M_{BH}$ in Equation (\ref{eq-dcycle}) refer to
  the mass of the SMBH at redshift $z$.

  If the distribution of $\tau$ does not evolve significantly
  over a time $t_{BL}$, then the integral in the numerator of Equation
  (\ref{eq-dcycle}) can be approximated as $p(\tau|M_{BH},BLQSO)
  t_{BL}$. Likewise, if $p(\tau|M_{BH},BLQSO)$ is approximately
  constant over $t_{BL}$, then the distribution of BLQSO initiation
  times will be similar to the distribution of BLQSOs at the time we
  observed them. This is a
  reasonable assumption so long as $t_{BL}$ is short compared to the
  timescale for a significant change in the process that initiates
  BLQSO activity (e.g., mergers and other fueling events). Making this
  assumption, and rearranging Equation (\ref{eq-dcycle}), it follows that
  \begin{equation}
    t_{BL} \approx \frac{\delta(M_{BH},z) \int_{0}^{t(z)}
      p(t'|M_{BH},BLQSO)\ dt'}{p(t(z)|M_{BH},BLQSO) Pr(BLQSO|M_{BH})}
    \label{eq-dcycle_approx}
  \end{equation}
  where $p(t(z)|M_{BH},BLQSO)$ is the probability distribution of
  BLQSOs as a function of cosmic age, $t(z)$. The term
  $p(t|M_{BH},BLQSO)$ is related to the BHMF for BLQSOs according to
  \begin{equation}
    p(t(z)|M_{BH},BLQSO) = \left|\frac{dt}{dz}\right|^{-1}
    \left[\frac{\phi(M_{BH},z)}{\int_{0}^{\infty} \phi(M_{BH},z)\
        dM_{BH}}\right] \label{eq-probt}.
  \end{equation}
  From Equation (\ref{eq-dcycle_approx}) it follows that for the
  special case where all SMBHs experience a BLQSO phase, and where
  BLQSOs initiation times are uniformly distributed over the age of
  the Universe, then $t_{BL} = \delta(M_{BH},z) t(z)$. This is the
  definition of a quasar `lifetime' commonly found in the literature;
  however, as the assumption of a uniform distribution of BLQSO
  initiation times is inconsistent with the BLQSO BHMF or luminosity
  function, $t_{BL}$ will in general not equal $\delta(M_{BH},z)
  t(z)$ \citep[see also][]{notlightbulbs}.

\begin{figure}
  \begin{center}
    \scalebox{0.7}{\rotatebox{90}{\plotone{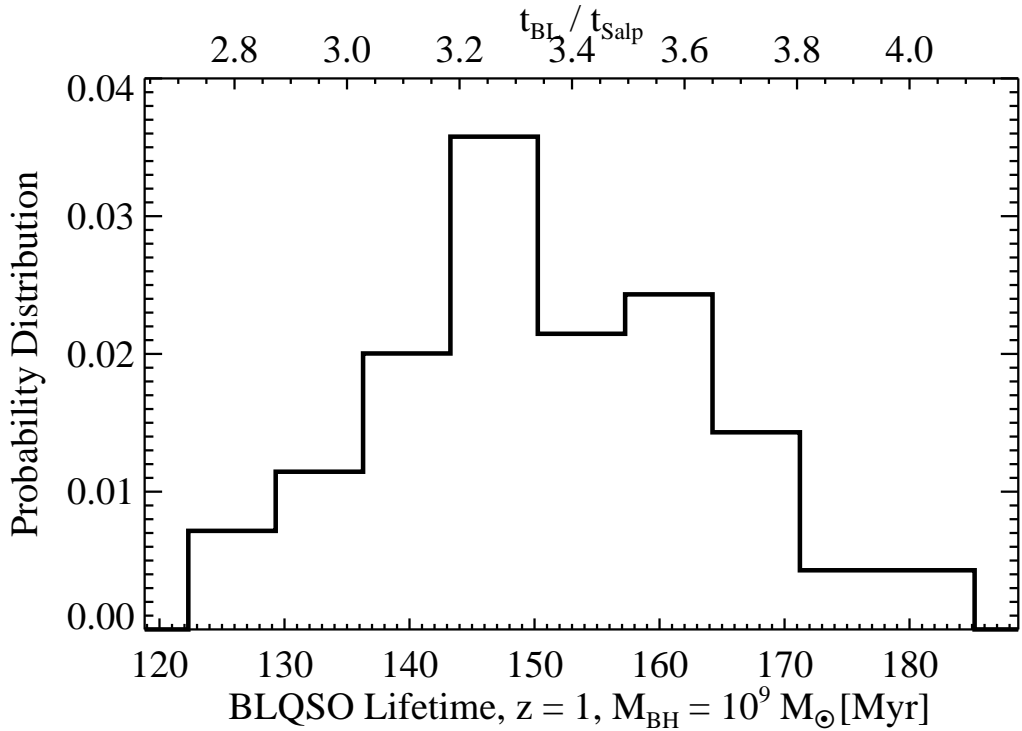}}}
    \caption{Probability distribution, given the observed data and
      assumptions of our analysis, of the lower bound on the
      BLQSO lifetime (Eq.[\ref{eq-dcycle_approx}]) calculated at $z =
      1$ for $M_{BH} = 10^9 M_{\odot}$. The upper axis gives the
      lifetime in terms of the Salpeter timescale, which
      is $t_{Salp} = 4.3 \times 10^7$ yrs for a radiative efficiency
      of $\epsilon_r = 0.1$. The lower-bound on the BLQSO lifetime is
      equal to the BLQSO lifetime if (1) all SMBHs of $M_{BH} = 10^9
      M_{\odot}$ go through a BLQSO phase and (2) if the number
      density of these SMBHs at $z = 1$ is not significantly less than
      the local number density. There is evidence that the latter
      condition is true \citep[e.g.,][]{merloni08}, while the former
      condition is expected if the SMBH's growth is self-regulated
      \citep{stochacc}. \label{f-blqso_lifetime}}
  \end{center}
\end{figure}

  Because we have assumed that $t_{BL}$ is the same for all BLQSOs of
  a given mass, and therefore must be the same at all redshifts,
  Equation (\ref{eq-dcycle_approx}) may be calculated at any redshift. However,
  in reality we can only estimate a lower limit to $t_{BL}$. This is
  because we do not know the fraction of SMBHs that will experience a
  BLQSO phase, although if the SMBHs growth is self-regulated we might
  expect $Pr(BLQSO|M_{BH}) \approx 1$. However, if some SMBHs of mass
  $M_{BH}$ can never be seen as BLQSOs, possibly due to
  orientation-dependent obscuration, then $Pr(BLQSO|M_{BH}) < 1$. In
  addition, as discussed 
  above, we can only calculate a lower limit to $\delta(M_{BH},z)$ by
  comparing the BLQSO BHMF at redshift $z$ with the local BHMF of all
  SMBHs. Moreover, implicit in these calculations is the assumption
  that none of these quantities depend strongly enough on mass to vary
  significantly during the growth that occurs in the BLQSO phase. In
  Figure \ref{f-blqso_lifetime} we show the probability 
  distribution of the BLQSO age of SMBHs with $M_{BH} = 10^9
  M_{\odot}$, calculated from Equation (\ref{eq-dcycle_approx}) at $z
  = 1$. We calculate $t_{BL}$ at $M_{BH} = 10^9 M_{\odot}$ because our
  sample is reasonably complete at this mass, especially at this
  redshift. In addition, we chose $z 
  = 1$ because the local BHMF better approximates the $z = 1$ BHMF
  than the BHMF at $z > 1$, therefore giving the tightest lower bound
  on the duty cycle at $z = 1$. We estimate $t_{BL} = 150 \pm 15$ Myr as
  a lower bound on the age of the BLQSO phase for SMBHs of $M_{BH} =
  10^9 M_{\odot}$. This estimate is roughly consistent with other
  estimates of quasar lifetimes
  \citep[e.g.,][]{yu02,martini04,gonc08,notlightbulbs}.

  \subsection{Constraints on The Most Massive Black Hole in the Universe}

  \label{s-massive}

  The probability distribution for the maximum mass of a SMBH in a
  BLQSO may place important constraints on models of SMBH growth, and
  may be calculated directly from the BHMF. In the context of our
  work, we do not consider the maximum mass of a SMBH to be caused
  by a hard upper limit, above which it is impossible to make a more
  massive black hole, but rather the result of a finite number of
  black holes drawn from a mass function. The probability that the
  maximum mass of a SMBH in a sample of $N$ BLQSOs is less than
  ${\cal M}$ is simply given by the probability that all $N$ SMBHs
  have $M_{BH} < {\cal M}$:
  \begin{equation}
    Pr(M_{BH}^{Max} < {\cal M}|N) = \left[ Pr(M_{BH} < {\cal M})
    \right]^{N} \label{eq-cdf_maxmass}
  \end{equation}
  Note that $N$ is the total number of BLQSOs that could be observed in
  an all-sky survey with no flux limit; i.e., $N$ is the normalization
  of the BLQSO BHMF. The term $Pr(M_{BH} < {\cal M})$ is calculated
  from the BHMF as
  \begin{equation}
    Pr(M_{BH} < {\cal M}) = \frac{1}{N} \int_{0}^{\cal M} \int_{0}^{\infty}
    \phi(M_{BH},z) \left(\frac{dV}{dz}\right) \ dz\
    dM_{BH}. \label{eq-cdf_bhmf}
  \end{equation}
  The probability distribution of $M_{BH}^{Max}$ is then found by
  differentiating Equation (\ref{eq-cdf_maxmass}) with respect to
  ${\cal M}$ and evaluating the result at ${\cal M} = M_{BH}^{Max}$:
  \begin{equation}
    p(M_{BH}^{Max} | N) = \left[ Pr(M_{BH} < M_{BH}^{Max}) \right]^{N-1}
    \int_{0}^{\infty} \phi(M_{BH},z) \left(\frac{dV}{dz}\right) \ dz.
    \label{eq-prob_maxmass}
  \end{equation}
  The posterior probability distribution for $M_{BH}^{Max}$, given the
  observed data, can be calculated by averaging Equation
  (\ref{eq-prob_maxmass}) over the MCMC output.

\begin{figure}
  \begin{center}
    \includegraphics[scale=0.33,angle=90]{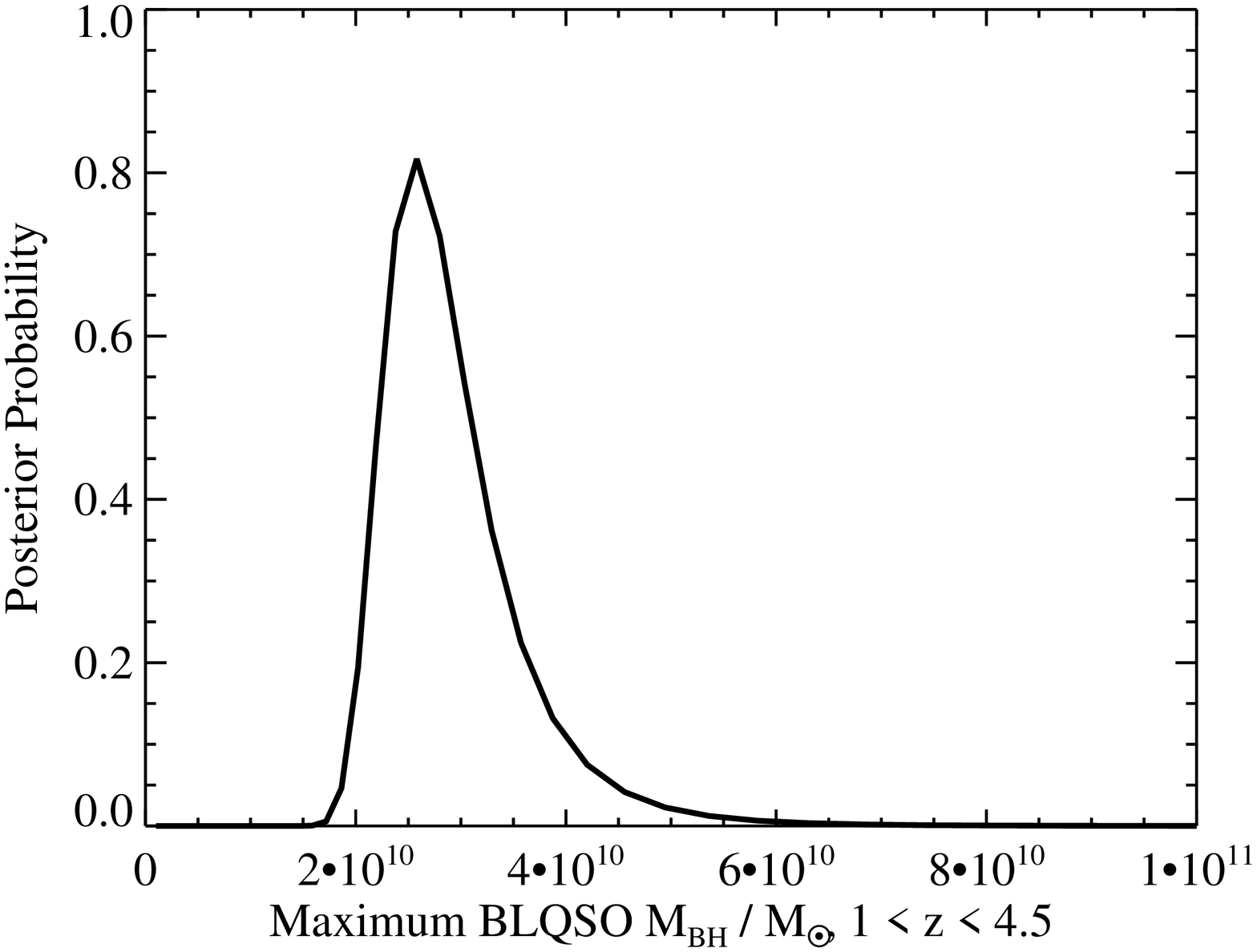}
    \includegraphics[scale=0.33,angle=90]{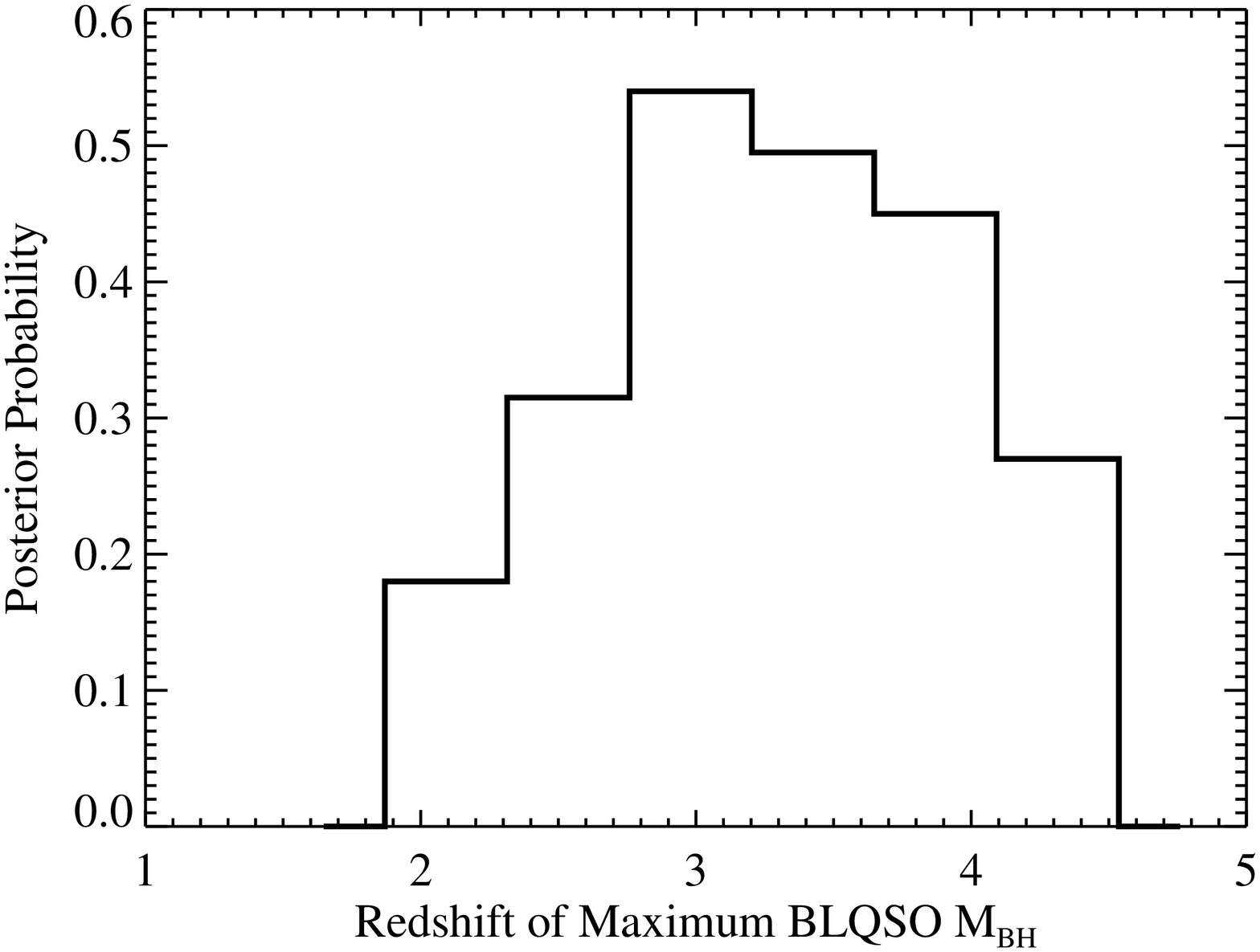}
    \caption{Probability distribution of $M_{BH}$ for the most massive
      SMBH that could be observed as a BLQSO at $1 < z < 4.5$ (left),
      and the probability distribution for the redshift that this SMBH
      would be seen as a BLQSO (right). Here, we do not interpret
      $M_{BH}^{Max}$ to represent a hard physical upper limit to
      $M_{BH}$, but rather it is the result of a finite number of
      black holes drawn from a mass function. The constraints on $M_{BH}^{Max}$ that we have
      obtained are consistent with previous observational work, but
      this is the first time that rigorous constraints on
      $M_{BH}^{Max}$ and its redshift have been obtained. The most
      likely value for $M_{BH}^{Max}$ is $M_{BH} \approx 2.5 \times 10^{10}
      M_{\odot}$, and this SMBH would likely be seen as a BLQSO
      at $z \gtrsim 2$. \label{f-maxmass}}
  \end{center}
\end{figure}

  In Figure \ref{f-maxmass} we show the posterior probability
  distribution of the maximum SMBH in a BLQSO at $1 < z < 4.5$. Here,
  $M_{BH}^{Max}$ should be interpreted as the maximum mass of a SMBH
  in a BLQSO that would be observed in an all-sky survey without a
  flux limit over the redshift interval $1 < z < 4.5$. Thus,
  $M_{BH}^{Max}$ is a lower bound on the mass of the most massive SMBH
  in the Universe. In addition, in Figure \ref{f-maxmass} we also show
  the probability distribution of the redshift at which the quasar
  with $M_{BH}^{Max}$ would be found. As can be seen from these
  figures, the maximum mass of a SMBH in a BLQSO at $1 < z < 4.5$ is
  $M_{BH}^{Max} \sim 3 \times 10^{10} M_{\odot}$. The probability
  distribution for the redshift of $M_{BH}^{Max}$ is rather broad, but
  we constrain the redshift for the BLQSO hosting this SMBH to be $z \gtrsim
  2$. 

  Our results are in agreement with what others have found using
  samples of broad line mass estimates
  \citep[e.g.,][]{vest04,vest08,netzer07}, although
  \citet{labita09a,labita09b} find a smaller value of $M_{BH}^{Max}
  \sim 5 \times 10^9 M_{\odot}$. However, in the model of
  \citet{labita09a} $M_{BH}^{Max}$ is a
  parameter determining the shape of the mass function, and is not the
  actual maximum black hole mass of a sample of objects drawn from the
  distribution of black hole mass; i.e., there is nothing in the
  statistical model of \citet{labita09a} that prevents objects with
  $M_{BH} > M_{BH}^{Max}$. As such, the actual realized maximum mass
  in a large sample of objects will be greater than the value of
  $M_{BH}^{Max}$ estimated using the model of \citet{labita09a}, as we have
  found. Considering this, our results are consistent with those of \citet{labita09a,labita09b}.

\begin{figure}
  \begin{center}
    \scalebox{0.7}{\rotatebox{90}{\plotone{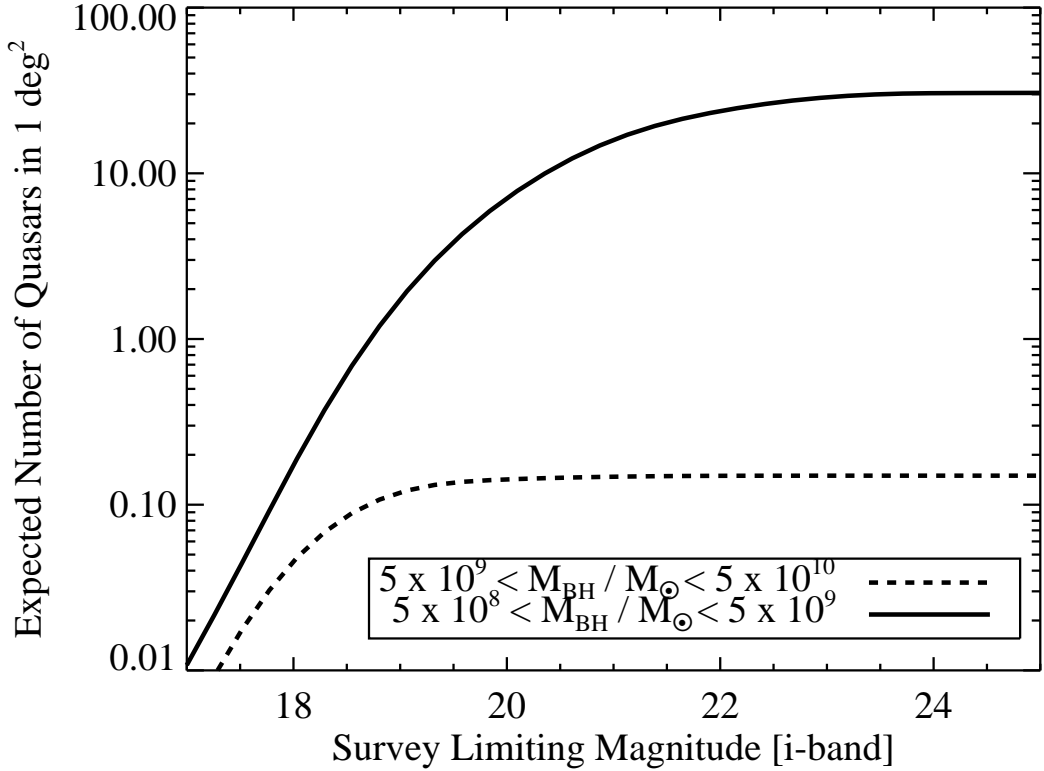}}}
    \caption{Expected number counts of BLQSOs at $1 < z < 4.5$ with
      black hole masses $5 \times 10^8 < M_{BH} / M_{\odot} < 5 \times
      10^9$ (solid line) and $5 \times 10^9 < M_{BH} / M_{\odot} < 5
      \times 10^{10}$ (dashed line) in a 1 ${\rm deg}^2$ survey as a function
      of limiting magnitude. One would need to search an angular area
      $\Omega > 10\ {\rm deg^2}$ in order to expect to find at least
      one BLQSO with $M_{BH} \sim 10^{10} M_{\odot}$. In addition, one
      does not become complete at $M_{BH} \sim 5 \times 10^8
      M_{\odot}$ until $i \sim 23$. \label{f-ncounts}}
    \end{center}
\end{figure}

  These highly massive black holes represent the extremes of black hole
  growth, and thus are important for constraining models of black hole
  growth. Furthermore, these massive black holes offer the best chance of
  probing the faint end of the BLQSO Eddington ratio
  distribution. Therefore, it is of use to investigate how large and
  deep a survey must be in order to find an useful number of these
  objects. In Figure \ref{f-ncounts} we show the expected number of high
  mass SMBHs in BLQSOs at $1 < z < 4.5$ in a $1\ {\rm deg}^2$ survey as a
  function of limiting $i$-magnitude, assuming our best-fit statistical
  model. We stress that these are the expected number counts for the
  true black hole masses, and not the mass estimates. Because the
  errors in the mass estimates will scatter more sources into higher
  mass bins than into lower mass bins, the number of sources with estimated mass
  in a mass bin will be larger than the actual number of sources.

  Any survey with a flux limit of $i < 19$ should be able to
  detect BLQSOs with $M_{BH} \gtrsim 5 \times 10^9 M_{\odot}$; however,
  the survey must have an area of $\Omega \gtrsim 10\ {\rm deg}^2$ to expect to
  detect at least one of these objects. Similarly, in order to get a
  large number of BLQSOs with $M_{BH} \gtrsim 5 \times 10^8 M_{\odot}$,
  one needs a deep, large area spectroscopic survey. For example, the
  BLQSO spectroscopic samples from the COSMOS survey \citep{cosmos}
  cover $\approx 2\ {\rm deg}^2$ at $i < 24$ \citep{trump07} and $i <
  22.5$ for zCOSMOS \citep{merloni10}, and, ignoring redshift
  incompleteness, are therefore expected to contain $\sim 60$ BLQSOs at
  $1 < z < 4.5$ with masses $5 \times 10^8 < M_{BH} / M_{\odot} < 5
  \times 10^9$, but none with masses $M_{BH} > 5 \times 10^9 M_{\odot}$.
  
  \subsection{Implied Eddington Ratio Distributions}

  \label{s-eddrat}

  As discussed in \S~\ref{s-statmod2}, we also estimate the
  distribution of the ratio of the Eddington ratio to the bolometric
  correction at 1350\AA, $\Gamma_{Edd} / C_{1350}$. Under the
  assumption that $\Gamma_{Edd} / C_{1350}$ follows a mixture of log-normal
  distributions, with mean values that depends linearly on $\log
  M_{BH}$, and that $C_{1350} = 4.3$ for all sources \citep{vest09}, our model
  implies that the geometric mean Eddington ratio increases with
  $M_{BH}$ according to:
  \begin{equation}
    \left\langle \frac{L}{L_{Edd}} \right\rangle_{\rm geo} = 0.12 \pm
    0.01 \left(\frac{M_{BH}}{M_{\odot}}\right)^{0.48 \pm 0.04}.
    \label{eq-gammean}
  \end{equation}
  The dispersion in $L / L_{Edd}$ decreases with increasing $M_{BH}$,
  having a value of $\sim 0.4$ dex at $M_{BH} \sim 10^8 M_{\odot}$, and
  decreasing to $\sim 0.3$ dex at $M_{BH} \sim 10^9 M_{\odot}$. Both
  the dispersion and slope are smaller than what was found by KVF09, who
  analyzed a sample of BLQSOs from the Bright Quasar Survey at $z <
  0.5$, suggesting possible evolution in the Eddington ratio
  distribution. However, the statistical significance of a difference
  in the slopes is marginal at best, as the differences are only significant at $\sim
  2\sigma$.

\begin{figure}
  \begin{center}
    \scalebox{0.7}{\rotatebox{90}{\plotone{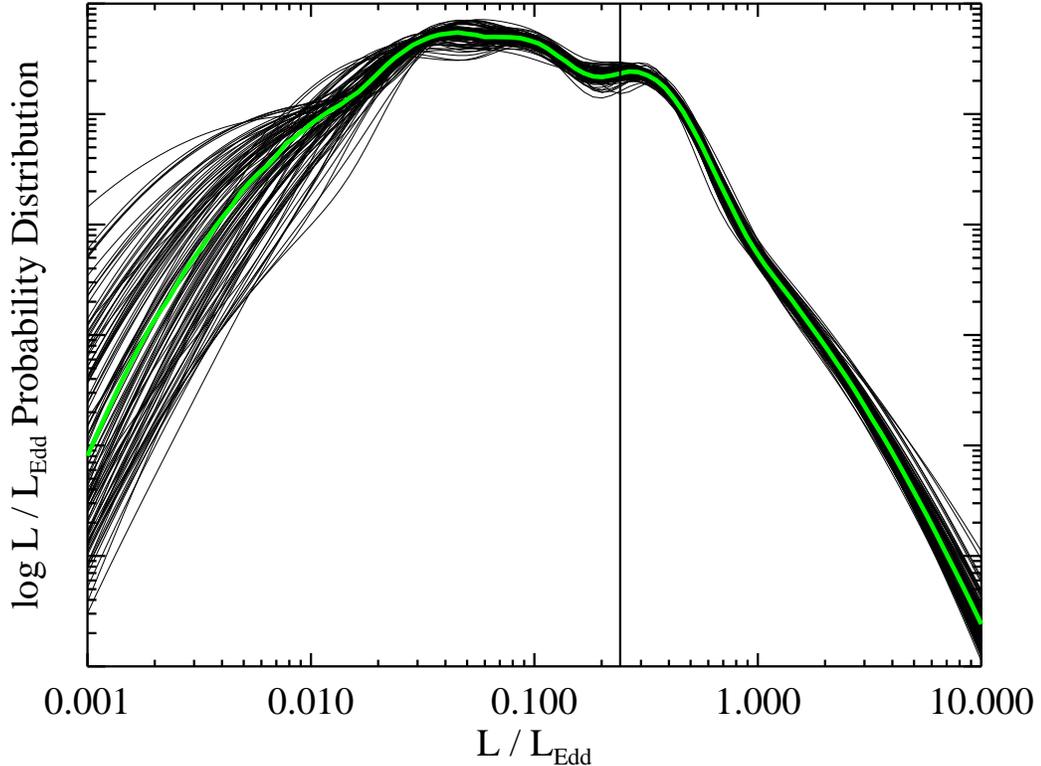}}}
    \caption{Inferred Eddington ratio distribution, assuming the
      mixture of log-normals form of Equation (\ref{eq-problm}) and a bolometric
      correction of $C_{1350} = 4.3$. The vertical line marks the
      $10\%$ completeness limit for the SDSS DR3 sample at $z = 1$;
      Figure \ref{f-completeness} shows that the $10\%$ completeness
      limits are similar for $z \sim 1, 3,$ and $4$, but shallower for
      $z \sim 2$. Our inferred Eddington ratio distribution peaks near $L / L_{Edd}
      \sim 0.05$ and has dispersion of $\sim 0.4$ 
      dex, although the peak fall below the $10\%$ completeness
      limit and should be interpreted with caution. Our inferred Eddington ratio distribution is shifted 
      toward lower values of $L / L_{Edd}$ and has a higher
      dispersion than previous work, which did not correct for
      incompleteness. Our estimated Eddington ratio distribution
      suggests that most SMBHs in BLQSOs are not radiating near their
      Eddington limit. \label{f-eratdists}}
  \end{center}
\end{figure}

  In Figure \ref{f-eratdists} we show the implied distribution of
  BLQSO Eddington ratios assuming a constant bolometric correction of
  $C_{1350} = 4.3$ \citep{vest09}. As can be seen, the estimated
  distribution of quasar Eddington ratios peaks at $L / L_{Edd} \sim
  0.05$. However we note that the location of the Eddington ratio peak falls below the
  $10\%$ completeness limit of the SDSS, and thus the exact location
  of the peak is highly uncertain. Our inferred Eddington ratio distribution is broader and
  shifted towards smaller values of $L / L_{Edd}$ than what has been
  found in previous work, which were not able to fully correct for incompleteness in
  black hole mass and Eddington ratio
  \citep[e.g.,][]{mclure04,vest04,koll06}. However, it is consistent
  with what \citet{shen08} found using a similar approach to ours. The
  major differences between our work and that of \citet{shen08} is that
  they assume a power-law distribution of $M_{BH}$, and they constrain
  the BHMF and Eddington ratio distribution by visually matching the
  model distributions to the observed distributions. Our results, as
  well as the results of \citet{shen08}, show that the narrow
  distribution in quasar Eddington ratios seen in previous work, and
  peaking at $L / L_{Edd} \sim 0.25$, is due to uncorrected
  incompleteness. Indeed, deeper samples of BLQSOs have observed
  Eddington ratio distributions that are broader and peak at lower
  values \citep{gav08,trump09}. Therefore, we conclude that most BLQSOs
  are not radiating at or near the Eddington limit, and that there is a
  large dispersion in Eddington ratio for BLQSOs.

  Recently, \citet{stein10a} have analyzed the SDSS DR5
  sample of mass estimates calculated by \citet{shen08}, and concluded
  that there is a dirth of objects at high Eddington ratio, and that
  there is a redshift-dependent systematic decrease in Eddington ratio with increasing
  black hole mass for BLQSOs in the high $L / L_{Edd}$ tail of
  the distribution. They have called this effect a sub-Eddington
  boundary. While we also find evidence that BLQSOs radiating near the
  Eddington limit are rare, as did KVF09, the dependence on black hole
  mass is seemingly in contrast to what we find here, in that we find
  that the average Eddington ratio increases with $M_{BH}$, assuming a
  constant bolometeric correction. The most important difference
  between their work and ours is the fact that
  \citet{stein10a} analyze seperate redshift bins, while we do not
  model a redshift dependence in the Eddington ratio
  distribution. \citet{stein10a} did not perform any calculations for
  the whole sample, averaged over all redshifts, making a more direct comparision
  difficult, and thus it is unclear how discrepant our conclusions are.

  Another potential contributor to this
  discrepancy is the handling of the 
  statistical error in the mass estimates. Statistical errors (e.g.,
  measurement error) have the effect of flattening slopes and
  correlations when one analyzes the quantity contaminated by the error
  \citep[e.g.,][]{akritas96,kelly07a,kelly07b}, due to the fact that the
  distribution of the estimated quantity is a biased estimate of the
  distribution of the quantity of interest. Because the mass estimates
  are contaminated by statistical error, the dependence of luminosity on
  the mass estimates will be flatter than the intrinsic dependence of
  luminosity on $M_{BH}$. In particular, the statistical error will
  cause some masses to be underestimated, and some to be
  overestimated. The objects which appear to have the highest masses at
  a given luminosity will have the highest overestimates, and thus will
  have their Eddington ratios underestimated. As a result, there will be
  an apparent dirth of high Eddington ratio objects at the highest
  masses. Similarly, there will be an apparent excess of high Eddington
  ratio objects in the low mass bins. Our method estimates the intrinsic
  dependence of luminosity on $M_{BH}$ by, in a sense, `deconvolving'
  the distribution of the mass estimates and luminosity with the
  distribution of the error in the mass estimates; this is also why we
  find a somewhat steeper decline in the mass function at the highest
  masses, as compared to the estimate reported by \citet{vest08}. 

  Although the statistical errors in the mass estimates can have the
  affect described above, possibly contributing to the different conclusions, it
  is unlikely that the statistical errors alone
  are sufficient to account for the discrepancy, and other
  possibilities include differences in the methods and scaling
  relationship 
  employed for estimating the masses, and other differences in the
  methods of data analysis employed. However, further investigation
  is beyond the scope of our work. In 
  addition, we note that our samples only overlap at $1 < z <
  2$\footnote[10]{\citet{stein10a} also included quasars at
    $z > 2$ in some of their plots, but their calculations and conclusions were based on
    quasars at $z < 2$ because of concerns regarding the validity of using C IV in the mass
    estimates.}, and we cannot compare with the low redshift results of
  \citet{stein10a}, where they find the greatest evidence for these
  effects.
  
  Although we have used a flexible form for $p(L|M_{BH})$, our
  estimated distribution for $L / L_{Edd}$ may be affected by
  significant systematics, as it relies on the assumption of a constant
  bolometric correction. There is evidence that the bolometric
  correction depends on both Eddington ratio \citep{vasud07,young10} and
  black hole mass \citep{kelly08}, and the increase of $L / L_{Edd}$
  with $M_{BH}$ we find may instead be a reflection of a decrease in the
  bolometeric correction with $M_{BH}$. Indeed, \citet{kelly08} find
  that the ratio of optical/UV flux to X-ray flux increases with
  $M_{BH}$, implying that the bolometeric correction to the UV flux
  decreases with $M_{BH}$, and therefore we would infer an increase in
  $L / L_{Edd}$ with $M_{BH}$ assuming a constant bolometeric
  correction, as we do. In addition to these issues, we also note that
  the SDSS is at least partially incomplete at all Eddington ratios at
  $z > 1$ \footnote[11]{As discusses earlier, this is merely a reflection of the fact that the mass
    function increases steeply toward lower masses, which are hard to
    detect even if these low mass SMBHs are radiating near the
    Eddington limit}, as shown in Figure
  \ref{f-completeness}, and further work is needed using deeper surveys
  to confirm these results.

  \section{DISCUSSION}

  \subsection{Connection with Quasar Lightcurves}

  \label{s-lcurves}

  A significant amount of recent work has suggested that quasar
feedback regulates the growth of SMBHs and affects the large-scale
evolution of its host galaxy. Understanding the quasar lightcurve is
thus of fundamental importance for understanding black hole growth and
feedback, as well as placing constraints on the physics of quasar
accretion flows. The distribution of luminosity at a given black hole
mass can be related to the quasar lightcurve, and therefore one can
use the estimated distribution of luminosity at a given black hole
mass to place some empirical constraints on models for quasar
lightcurves. We address this in the following section, and argue that
our estimated Eddington ratio distribution is consistent with models
where the BLQSO phase represents the final stages of a SMBH's growth,
in which the QSO lightcurve is expected to decay until the object no
longer appears as a BLQSO.

  For a given fueling episode, denote the quasar lightcurve as $L(t)$,
  and the mass fueling rate to the SMBH as a function of time as
  $\dot{M}(t)$. The fueling rate, $\dot{M}(t)$, gives the rate at
  which matter is externally supplied to the BLQSO accretion disk. The
  probability distribution of quasar luminosity at a given $M_{BH},
  \dot{M},$ and $t$ is $p(L|M_{BH},\dot{M},t)$, and the probability
  distribution of the fueling rate at a given $M_{BH}$ and $t$ is
  $p(\dot{M}|M_{BH},t)$. The distribution $p(L|M_{BH},\dot{M},t)$ is
  determined by the physics of the accretion disk, since the quasar
  luminosity is produced by viscous stresses in the accretion disk. For example, \citet{siem97} 
  calculate $p(L|M_{BH},\dot{M},t)$ implied by the model of \citet{siem96}
  for a thermal-viscous accretion disk stability. The distribution
  $p(\dot{M}|M_{BH},t)$, on the other hand, depends on the physics
  and stochastic nature of the fueling mechanism and quasar
  feedback. 

  Because the quasar luminosity is determined by the physics
  of the accretion disk, it is
  reasonable to assume that given $M_{BH}$ and $\dot{M}$, $L$ is
  independent of time. This does not imply that the quasar lightcurve
  does not vary with time, as it certainly does, but rather that
  knowing the value of $t$ does not give us any additional information
  on the value of $L$ when we already know $M_{BH}$ and $\dot{M}$ at a
  given $t$. We therefore drop the explicit dependence of
  $p(L|M_{BH},\dot{M},t)$ on time.

  The distribution of luminosity at a given $M_{BH}$ is calculated as
  \begin{equation}
    p(L|M_{BH}) =  \int_{0}^{\infty} p(L|\dot{M},M_{BH})
    \left[\int_{t_0}^{t_o+t_{BL}} p(\dot{M}|t,M_{BH}) p(t|M_{BH})\ dt
    \right] \ d\dot{M},
    \label{eq-timeint}
  \end{equation}
  where $p(t|M_{BH})$ is the probability of seeing a BLQSO at time
  $t$, given $M_{BH}$, and $t_{BL}$ is the length of time that a
  quasar would be classified as a BLQSO. Here we have defined the
  broad line phase of quasar activity to start at $t=t_0$. The term
  $p(t|M_{BH})$ is related to the model for black hole growth, since
  $p(t|M_{BH}) \propto p(M_{BH}|t) p(t)$. Because we observe quasars
  randomly during their BLQSO phase, $p(t)$ is uniform and $p(t|M_{BH})
  \propto p(M_{BH}|t)$. In general, BLQSOs with larger values
  of $M_{BH}$ are more likely to be seen later in their lifetime, i.e.,
  at larger values of $t$.

  As an alternative to Equation (\ref{eq-timeint}), 
  $p(L|M_{BH})$ may be directly calculated from the amount of time
  that a BLQSO spends above a given luminosity, as a function of
  $M_{BH}$. In this case, the term $p(L|M_{BH})$ is directly related
  to the `luminosity-dependent' lifetime interpretation advocated by
  \citet{lumdep_lifetime1,lumdep_lifetime2}, where the quasar
  `lifetime' is the amount of time that a quasar spends above a given
  luminosity. Assume that $p(t|M_{BH}) \propto 1$ for BLQSOs, and denote
  $T(L|M_{BH})$ to be the amount of time a BLQSO spends above a
  luminosity $L$, as a function of $M_{BH}$. Then
  \begin{equation}
    p(L|M_{BH}) \propto \left| \frac{dT(L|M_{BH})}{dL} \right|.
    \label{eq-lumdep_lifetime}
  \end{equation}
  For example, \citet{notlightbulbs} suggest that $p(L|M_{BH})$ can be
  well approximated as a Schechter function, which generalizes the
  distributions implied by several common models of quasar lightcurves
  often found in the literature.

  \subsubsection{Effects of Evolution in the Quasar Fueling Rate}

  \label{s-fuelrate}

  Recent work on quasar fueling and feedback suggests that the BLQSO
  phase occurs at the end of the fueling event, during which
  feedback energy from the AGN is able to unbind the surrounding gas,
  removing the obscuring material and 
  halting the black hole's growth. Within the context
  of this model, the SMBH does not experience a large fractional
  increase in mass beyond the value
  of $M_{BH}$ that is observed during its time as a BLQSO. Therefore,
  we approximate $p(t|M_{BH})$ as being uniform and independent of
  $M_{BH}$, i.e., $p(t|M_{BH}) = 1 / t_{BL}$. In addition, this model predicts that during this so-called
  `blow-out' phase, the fuel supply is set by the evolution of the
  blastwave caused by the injection of energy from 
  the SMBH, and is expected to be of the form $\dot{M}(t) \propto
  t^{-\beta}$ \citep{stochacc,faintend}. Alternatively, if the fuel
  supply is suddenly removed, then the accretion rate during the
  broad line phase is due to evolution of the viscous accretion
  disk. The fueling rate also has a power-law form under viscous
  evolution of the disk, but with a different value for $\beta$
  \citep{cann90,pringle91}. Under 
  these models, it is reasonable to approximate the fuel supply as a
  deterministic process, $\dot{M}(t) \propto t^{-\beta}$. If the evolution
  of the fuel supply is a deterministic and one-to-one process, and
  assuming that $p(t|M_{BH}) \propto 1 / t_{BL}$, then
  $p(\dot{M}|M_{BH})$ is 
  \begin{equation}
    p(\dot{M}|M_{BH}) = \frac{1}{t_{BL}} \frac{dt(\dot{M},M_{BH})}{d\dot{M}},
    \label{eq-lcurve_determ}
  \end{equation}
  where $t(\dot{M},M_{BH})$ is the time implied by $\dot{M}(t,M_{BH})$, i.e.,
  $t(\dot{M},M_{BH})$ is the inverse function of $\dot{M}(t,M_{BH})$. If the
  quasar lightcurve is a many-to-one function, then the right side of
  Equation (\ref{eq-lcurve_determ}) is replaced by a sum of
  derivatives, as in Equation (15) of \citet{yu04}. Inserting Equation
  (\ref{eq-lcurve_determ}) into Equation (\ref{eq-timeint}) and
  dropping the time integral gives\footnote[12]{This follows from
    $p(L|M_{BH}) = \int p(L|\dot{M},M_{BH}) p(\dot{M}|M_{BH})\
    d\dot{M}$. Alternatively, we could have arrived at
    Eq.[\ref{eq-timeint2}] by taking $p(\dot{M}|M_{BH},t)$ to be a
    delta function centered at the value of $\dot{M}$ determined by
    $t$, and directly doing the integration in Eq.[\ref{eq-timeint}].}
 \begin{equation}
    p(L|M_{BH}) =  \frac{1}{t_{BL}} \int_{0}^{\infty}
    p(L|\dot{M},M_{BH}) \frac{dt(\dot{M},M_{BH})}{d\dot{M}}\ d\dot{M}.
    \label{eq-timeint2}
  \end{equation}

  For models where $\dot{M}(t) \propto t^{-\beta}$, a power-law distribution of
  fueling rates follows from Equation (\ref{eq-lcurve_determ}),
  $p(\dot{M}|M_{BH}) \propto \dot{M}^{-(1 + 1/\beta)}$ \citep[see
    also Eq. 43 in][]{yu08}. Therefore, assuming $\dot{M}(t) \propto
    t^{-\beta}$ gives
 \begin{equation}
    p(L|M_{BH}) \propto \int_{0}^{\infty} p(L|\dot{M},M_{BH})
    \dot{M}^{-(1 + 1/\beta)}\ d\dot{M}.
    \label{eq-timeint_plaw}
  \end{equation}
  The feedback-driven model of \citet{stochacc} predicts a value of
  $\beta \sim 2$, while the disk evolution model 
  predicts $\beta \sim 1.2$ \citep{cann90,yu05,king07}, implying that
  $p(\dot{M}|M_{BH}) \propto \dot{M}^{-1.5}$ or $p(\dot{M}|M_{BH}) \propto
  \dot{M}^{-1.67}$, respectively.

  \subsubsection{Effects of Time-Dependent Accretion Disks}

  \label{s-diskevol}

  It is apparent from Equation (\ref{eq-timeint_plaw}) that the effect
  of time-dependent accretion disk behavior is to create a
  distribution of luminosities at a given $M_{BH}$ that is flatter
  than that expected from a simple power-law decay in the fueling
  rate.  While the BLQSO fueling rate may be approximated as a
  deterministic process, quasar lightcurves are observed to exhibit
  aperiodic and stochastic variations across all wavelengths
  \citep[for a review see][]{ulrich97}. Therefore, the quasar
  lightcurve will be stochastic, i.e., the value of the luminosity is
  not completely determined by $M_{BH},$ and $\dot{M}$. From Equation
  (\ref{eq-timeint_plaw}) it can be observed that if we assume a
  deterministic relationship between $L$ and $\dot{M}$, i.e., a
  time-steady accretion disk, as much previous work on SMBH growth and
  feedback has assumed, then $L \propto \dot{M}$ and the quasar
  lightcurve simply traces the fueling rate evolution. In this case,
  $p(L|M_{BH},\dot{M})$ is a delta function and $p(L|M_{BH}) \propto
  L^{-(1 + 1/\beta)}$. However, the time-dependent and stochastic
  nature of the accretion disk emission broadens the observed
  luminosity function beyond that implied solely from the BLQSO
  fueling evolution.

  Although quasars are observed to vary at all wavelengths, we focus
  our remaining discussion on optical variability, since we are
  interested in the distribution of optical luminosities in this work.
  \citet{kelly09} found that quasar optical lightcurves on timescales
  $\lesssim 7$ yrs in the rest frame of the quasar are well-described
  by a Gaussian process on the 
  logarithmic scale \citep[see also][]{koz10}, with characteristic timescale consistent with
  accretion disk thermal time scales. They suggested that quasar
  variability on these time scales is due to a turbulent magnetic
  field in the accretion disk, which drives changes in the radiation energy density of the
  disk, as also seen in 3-dimensional magneto-hydrodynamic simulations
  of radiation-pressure dominated accretion disks
  \citep{hirose08}. Extrapolation of their best-fit lightcurves imply
  that flux variations for this particular process have a standard
  deviation $\sim 0.1$ dex on timescales long compared to the disk
  thermal time scale, independent of $M_{BH}$. If this is the only
  source of quasar optical variability, then $p(L|M_{BH},\dot{M})$ is
  a gaussian distribution centered at $L_{opt} = C_{opt} \epsilon_r
  \dot{M} c^2$ with standard deviation $\sim 0.1$ dex. Here $C_{opt}$
  is the bolometric correction to the optical luminosity (which likely
  depends on $M_{BH}$ and $\dot{M}$), $\epsilon_r$ is the radiative
  efficiency, and $c$ is the speed of light. This process only
  produces a small amount of broadening in the luminosity
  distribution, relative to that implied solely from the BLQSO fueling
  evolution. Therefore, $p(L|M_{BH}) \propto L^{-(1 + 1/\beta)}$ is a
  reasonable approximation when the fuel rate declines as a power-law,
  so long as there are no other variability components in the
  accretion disk beyond that observed by \citet{kelly09}. However,
  this does not avoid the issue of a non-constant bolometric correction.
 
  Variability on longer time scales may be driven by accretion disk
  instabilities. Accretion disks based on the standard
  $\alpha$-prescription for viscosity \citep{shak73} are subject to a
  number of accretion disk instabilities that can potentially have a
  significant effect on the quasar lightcurve. At high accretion rates
  ($\dot{m} \gtrsim 0.025$) a radiation-pressure instability may
  operate on time scales $\gtrsim 10^4$ yrs \citep{czerny09}, while at
  lower accretion rates a thermal-viscous ionization instability
  \citep[e.g.,][]{lin86,siem96,menou01,janiuk04} may operate on time
  scales $\gtrsim 10^6$ yrs. The ionization instability has been
  invoked to explain the outbursts seen in dwarf novae and soft X-ray
  transients \citep[for a review see][]{lasota01}, and the radiation
  pressure instability has been invoked to explain the outbursts seen
  in the microquasar GRS 1915+105
  \citep[e.g.,][]{nayakshin00,janiuk00}. Moreover, \citet{good04}
  suggest that stars may also form in AGN accretion disks due to
  fragmentation in the outer edges of the disk. While it is apparent
  that accretion disks around galactic sources exhibit instabilities, it is
  currently unclear if and how these instabilities operate in AGN, as
  the time scales involved are too long to observe transitions between
  quiescence and outburst. Furthermore, theoretical predictions of
  lightcurves based on these instabilities vary, and the existence and
  importance of disk stabilities can depend on how the viscosity is
  parameterized \citep[e.g.,][]{stella84,szusz90,merloni03}, how much
  of the accretion energy is dissipated in a hot corona
  \citep[e.g.,][]{svensson94}, or if there is an outflow or an
  advection dominated accretion flow (ADAF)
  \citep[e.g.,][]{hameury09}.

  The location in the disk where the instability occurs is important,
as $p(L|\dot{M},M_{BH})$ is with respect to the optical flux in our
model. Various instabilities are expected to operate in different
regions of the disc, sometimes in a stochastic manner, and thus may or
may not significantly affect the optical flux. The term
$p(L|\dot{M},M_{BH})$ is also conditional on the source being seen as
a BLQSO, and if the disk instabilities cause the ionizing continuum to
disappear during quiescence, thus causing the broad emission lines to disappear as well, then $p(L|\dot{M},M_{BH})$ is only with regard
to the distribution of optical luminosities during a disk
outburst. Considering these issues, and the current uncertainty
regarding the importance of disk instabilities, we consider it beyond
scope of this paper to fully investigate the effects of time-dependent
behavior in the accretion disk. However, in light of this discussion,
disk instabilities can potentially have a significant effect on the
distribution of luminosity at a given $M_{BH}$ \citep{siem97}. Thus, the distribution of luminosity at
a given $M_{BH}$ is likely altered beyond a simple power-law expected
if the optical flux is trivially related to the external fueling rate.

  In spite of these issues related to the detailed physics of the
accretion disk, our estimated $p(L|M_{BH})$ is qualitatively
consistent with models which predict the BLQSO phase to occur while
the quasar accretion rate is decaying, as we find that high Eddington
ratio objects are rare, and that the number density of BLQSOs 
increases steeply toward lower values of $L / L_{Edd}$. In addition,
although our estimated Eddington ratio distribution continues to rise
toward lower $L / L_{Edd}$, our sample is too incomplete to rigorously probe the
low Eddington ratio region of the distribution, and our estimated
$L / L_{Edd}$ distribution in this region is heavily dependent on our
parameteric form used to extrapolate beyond $L / L_{Edd} \lesssim
0.1$. Thus, we cannot test if the Eddington ratio distribution
continues to rise until $L / L_{Edd} \sim 10^{-2}$, after which a
steep decay in $p(L / L_{Edd})$ might occur due to the disappearance
of the broad emission lines \citep[e.g.][]{churazov05,trump09}.

  \subsection{Implications for the Growth of Supermassive Black Holes}

  \label{s-bhgrowth}

  Our results in this work are broadly in agreement with recent
  observational and theoretical work suggesting that SMBH growth and
  spheroid formation have a common origin. In particular, models where
  mergers dominate black hole growth predict that major mergers of
  gas-rich galaxies initiate a burst of star formation
  \citep{mihos94a,mihos96}, during which the black hole undergoes
  Eddington-limited obscured growth. Eventually the black hole becomes
  massive enough for radiation-driven feedback to unbind the
  surrounding gas, halting the accretion flow and revealing the object
  as a BLQSO \citep[e.g.,][]{hopkins_long}. As the activity further
  declines, the remnant will redden and become quiescent, satisfying
  the black hole--host galaxy correlation and leaving a dense stellar
  remnant from the starburst \citep{mihos94b}. This dense stellar
  remnant is identifiable as a second component in the light profiles
  of elliptical galaxies \citep{hop08b,hop09c,hop09d}. Techniques
  based on the argument of \citet{soltan82} have concluded that the
  SMBH accretion rate density of the Universe peaks at $z \sim 2$
  \citep[e.g.,][]{marconi04,merloni08,shank09}, in agreement with the
  observed peak in the SMBH luminosity density of the Universe at $z \sim
  2$ \citep[e.g.,][]{wolf03,hopkins07}. We have found that the peak in
  the BLQSO cosmic mass density also occurs at $z \sim 2$, in broad
  agreement with models where the SMBH undergoes Eddington-limited
  growth up to the BLQSO phase, at least on a cosmological level.

  In addition, our results suggest that a lower bound on the typical
  length of time for which a massive ($M_{BH} \sim 10^9 M_{\odot}$)
  SMBH could be seen as a BLQSO during a single BLQSO episode is $t_{BL} \gtrsim 120$ Myr, i.e.,
  a few Salpeter times. If all $M_{BH} \sim 10^9 M_{\odot}$ SMBHs
  go through a BLQSO phase, and if the number density of SMBHs with
  $M_{BH} \sim 10^9 M_{\odot}$ does not significantly increase from $z
  = 1$ to $z = 0$, then Equation (\ref{eq-dcycle_approx}) is no longer
  a lower bound and $t_{BL} \sim 150$ Myr represents an estimate of the
  lifetime of a single BLQSO phase. Indeed, if the SMBH's growth is
  self-regulated then there is good reason to assume that most SMBHs go
  through a BLQSO phase at the end of their growth
  \citep{stochacc}, as the obscured/unobscured dichotomy represents an
  evolutionary sequence, at least at $z > 1$. Furthermore, recent work employing the argument of
  \citet{soltan82} suggests that the number density of $M_{BH} \sim
  10^9 M_{\odot}$ SMBHs does not significantly increase from $z = 1$
  to $z = 0$ \citep{merloni08}. These two considerations imply that
  $t_{BL} \sim 150$ Myr should represent a reasonable estimate of the
  length of the BLQSO phase. 

  Our estimated BLQSO lifetime of $t_{BL} \sim 150$ Myr is longer than
  the value of $t_{BL} \sim 10$--$20$ Myr predicted by
  \citet{hopkins_long}. Part of this discrepancy may be due to the
  fact that they defined a BLQSO as having a
  B-band luminosity brighter than some fraction of the host galaxy's,
  while we define a BLQSO as simply having broad emission lines along
  our line of sight. Therefore, if a
  BLQSO has a decaying lightcurve, as assumed in the model of
  \citet{hopkins_long}, then by our definition a SMBH may still be
  considered a BLQSO even after the B-band luminosity has fallen below
  its host's. This would imply that the predicted BLQSO lifetime under
  the definition of \citet{hopkins_long} would be shorter than our
  estimated value, as is the case. However, it is unclear if the
  lifetime of the BLQSO should be a factor of $\sim 10$ shorter under
  the definition of \citet{hopkins_long}. Unfortunately, we cannot
  invert our estimated Eddington ratio distribution to a quasar
  lightcurve to test this, as the inversion is not unique, and we
  would have to assume a distribution of host galaxy B-band luminosity
  during the BLQSO phase at $z \sim 1$, which is poorly
  constrained; as \citet{hopkins_long} point out, uncertainty in the
  ratio of host galaxy to quasar B-band luminosity also introduces
  considerable systematic uncertainty in their predicted BLQSO
  lifetime. Moreover, our estimated lifetime is subject to the
  assumptions outlined in \S~\ref{s-lifetime}, which may introduce
  additional systematic uncertainties of a factor of a
  few. Considering this, a more quantitative comparison between 
  our estimate and the prediction from \citet{hopkins_long} is difficult.

  Our estimated BLQSO BHMF suggests a typical value of $M_{BH} \sim
  10^8 M_{\odot}$ for BLQSOs.  Theoretical estimates of the mass
  distribution of SMBH seeds suggets that typical values for the seed
  mass should be $M_{BH} \sim 10^5$--$10^6 M_{\odot}$ at $z \sim 10$
  \citep{lodato07,pel07,vol08}, and grow to $M_{BH} \sim 10^8
  M_{\odot}$ by $z \sim 3$ \citep{vol09}, consistent with our
  results. If the black hole's growth is 
  Eddington-limited, then at least several Salpeter times are required
  to grow a seed black hole to $M_{BH} \sim 10^9 M_{\odot}$ from a
  $M_{BH} \sim 10^6 M_{\odot}$ seed, which is inconsistent with our
  estimated BLQSO lifetime. Furthermore, we find that most SMBHs in
  BLQSOs are not radiating near the Eddington limit, with a typical
  value being $L / L_{Edd} \sim 0.1$ at $M_{BH} \sim 10^9 M_{\odot}$,
  as we might expect if BLQSOs are 
  seen during a phase with a decaying fueling rate. Because most BLQSOs
  are radiating at considerably less than the Eddington limit, this
  therefore suggests that a factor of $\sim 10$ longer is needed to grow
  these SMBHs from seeds of $M_{BH} \sim 10^6 M_{\odot}$, i.e., a growth
  time scale of $\sim 70$ Salpeter times. This corresponds to the age
  of the universe at $z \sim 2$, implying that sources observed at
  $z > 2$ had to have been accreting at higher rates earlier in their
  growth. 

  An inferred growth time scale of $\sim 70$ Salpeter times is
  significantly longer than our estimated BLQSO lifetime at $z \sim 1$,
  implying one of a few explanations. First, it implies that if we assume that
  BLQSO SMBHs spend all of their growth as a BLQSO, then our assumption
  that all SMBHs go through a BLQSO phase along our line of sight is incorrect. In other words,
  this implies that some of the SMBH population that is growing at any
  redshift could never be observed by us to be a BLQSO at any
  time. Because the lifetime is related to what we can calculate from
  the BHMF by Equation (\ref{eq-dcycle_approx}), and if the SMBH spends
  all of its growth as a BLQSO, then an inferred SMBH growth time scale
  of $t_{BL} \sim 70 t_{salp}$ implies that only $\sim 4\%$ of SMBHs
  ever go through a phase where we could observe them as BLQSOs at any
  time. This is an unrealistically low number, and thus we conclude that $t_{BL}$ is
  shorter than the growth time. Furthermore, it also illustrates that
  even if we can never observe a large fraction of active SMBHs, say
  $\sim 50-75\%$, possibly due to orientation-dependent obscuration,
  then our estimated $t_{BL}$ is only underestimated by a factor of
  $\sim 2$--$4$.

  Another possibility is that SMBHs in BLQSOs at $z \sim 1$ built up
  their mass via numerous fueling events of length $t_{BL} \gtrsim
  150$ Myr. In this case one could grow a SMBH from a seed mass of
  $M_{BH} \sim 10^6 M_{\odot}$ at $z \sim 10$, say, to a mass of
  $M_{BH} \sim 10^9 M_{\odot}$ by $z \sim 1$, without the need for an
  earlier phase of obscured and accelerated growth. Similarly, we may
  have incorrectly assumed that the lifetime of the BLQSO phase is
  short compared to the time scale needed for any significant change
  in the BLQSO triggering time distribution, as this assumption was
  needed in order to use the observed distribution of BLQSOs as an
  estimate of their triggering distribution. This assumption may be
  incorrect if BLQSOs undergo a single long growth phase from $z \sim
  10$ to $z \sim 1$. However, this possibility only exists for $z
  \lesssim 1.7$, since at higher redshifts the growth time for objects
  radiating at $L / L_{Edd} \sim 0.1$ is longer
  than the lookback time to $z \sim 10$, and thus BLQSOs that are
  observed at $z \gtrsim 1.7$ would have had to experience an earlier
  phase of obscured growth at an enhanced accretion rate. Moreover, if
  SMBHs that are seen as BLQSOs at $z \sim 1$ with $M_{BH} \sim 10^9
  M_{\odot}$ are grown in a single long BLQSO episode, or numerous
  repeated ones of $t_{BL} \sim 150$ Myr, this would
  also imply that BLQSOs that are seen at $z \lesssim 1.7$ would have
  had a different fueling mechanism than BLQSOs that are seen at $z
  \gtrsim 1.7$, and it is unclear why this should be true. Therefore,
  we conclude that while part of the mass of SMBHs in BLQSOs may have
  been accumulated via multiple BLQSO episodes, it is unlikely that
  most of these SMBHs did not also experience a phase of obscured growth at an
  enhanced accretion rate.

  If most SMBHs go through a BLQSO phase along our line of sight at some point
  in their growth, the fact that our estimated BLQSO lifetime is short
  compared to the SMBH growth time implies that SMBHs spend a
  significant amount of their time growing in a non-BLQSO phase, such as
  an obscured phase. Note that this argument is unaffected by the fact
  that at any $z$ we miss a considerable fraction of the SMBH population
  that is growing (e.g., due to obscuration), so long as these missed SMBHs
  undergo a BLQSO episode at some point in their life. The conclusion
  that a significant amount of growth occurs in an earlier obscured
  phase was also reached by \citet{treist10}, and is expected from
  self-regulation models models where mergers or other triggering mechanisms fuel and initiate
  Eddington-limited accretion and obscured SMBH growth, until the SMBH
  becomes massive enough to unbind the ambient gas, revealing it as a
  BLQSO for $t_{BL} \sim 150$ Myr. Within this interpretation, the BLQSO
  black hole mass density shown in Figure \ref{f-massdensity} is
  proportional to the rate at which the relic SMBH mass increases as a
  function of redshift. Furthermore, considering that the growth time
  scale for a SMBH radiating at $L / L_{Edd} \sim 0.1$ to grow from
  $10^6 M_{\odot}$ to $10^9 M_{\odot}$ corresponds to the age of the
  universe at $z \sim 2$, we also conclude that SMBHs accrete at a
  significantly higher rate during the earlier obscured phase, as
  compared to the BLQSO phase.

  In this work we have found that the most massive SMBH that could be
  seen as a BLQSO is $M_{BH}^{Max} \sim 3 \times 10^{10} M_{\odot}$ and would
  most likely be observed at $z \gtrsim 2$.  These constraints on
  $M_{BH}^{Max}$ are consistent with recent cosmological simulations
  of SMBH growth, as well as expectations from black hole feedback
  models. Cosmological simulations that follow the growth of SMBHs in
  bright $z \sim 6$ quasars have been able to grow SMBHs to $M_{BH}
  \sim 10^9 M_{\odot}$ by $z \sim 4$ \citep[]{li07,dimatteo08} and
  $M_{BH} \sim 2 \times 10^{10} M_{\odot}$ by $z \sim 2$
  \citep{sijacki09}. Similarly, considerations based on quasar
  feedback and self-regulated SMBH growth, combined with the local
  distribution of bulge velocity dispersion, also suggest a value of
  $M^{Max}_{BH} \sim 10^{10} M_{\odot}$ \citep{priya09}.

  If the SMBHs growth is self-regulated, then the final mass of the
  SMBH after a fueling event is set by the binding energy of the bulge
  regardless of the fueling mechanism \citep{younger08}. In order to
  buildup a mass as large as $M_{BH}^{Max}$, multiple fueling episodes
  would likely be necessary, as is seen in cosmological simulations
  \citep[e.g.,][]{li07,dimatteo08,sijacki09}. However, it is unlikely
  that $M_{BH}^{Max}$ could be increased significantly beyond the
  observed value as additional fueling mechanisms would likely result
  in only a small increase in the binding energy of the
  bulge. Therefore, additional fueling episodes would immediately
  result in the SMBH unbinding the accreting gas, preventing
  significant additional growth. Indeed, if our estimated BLQSO
  lifetime of $t_{BL} \sim 150$ Myr is correct for SMBHs with $M_{BH}
  \sim 10^9 M_{\odot}$, and if the SMBH is radiating at a typical
  Eddington ratio of $L / L_{Edd} \sim 0.1$ during the BLQSO phase,
  then it would only accrete an additional $\sim 10^8 M_{\odot}$,
  assuming a radiative efficiency of $\epsilon_r = 0.1$. If the
  distribution of Eddington ratios is a power-law with
  $p(\Gamma_{Edd}) \propto \Gamma_{Edd}^{-1.5}$, as suggested by the
  discussion in \S~\ref{s-fuelrate}, then the typical Eddington ratio
  would be $L / L_{Edd} \lesssim 0.1$, suggesting even less
  growth during the BLQSO phase. Moreover, additional growth via black
  hole mergers, as might be expected from `dry' mergers of galaxies,
  is also unlikely to lead to significant additional growth, as the
  mass of the additional SMBH is likely to be negligible compared to
  $M_{BH}^{Max}$. And finally, additional growth through radiatively
  inefficient low-accretion rate modes does not contribute
  significantly to the black hole's final mass
  \citep{hnh06,cao07}. Therefore, our estimated value of $M_{BH}^{Max}
  \sim 3 \times 10^{10} M_{\odot}$ should be representative of the most massive
  SMBH for both active and inactive SMBHs.

  Our results are consistent with previous data-based work that has
  attempted to map the distribution and growth of SMBHs. Most previous
  observational work has mapped SMBH growth by using the
  $M_{BH}$--$\sigma$ relationship to infer the local BHMF, and then
  stepped backward in time using the quasar luminosity function to infer
  the contribution to the BHMF from accretion
  \citep[e.g.,][]{soltan82,yu02,shank04,marconi04,merloni08}. In
  addition, there have been attempts to predict the BHMF and its
  evolution for all SMBHs, or all active SMBHs
  \citep[e.g.,][]{vol03,sijacki07,hopkins_cosmo,dimatteo08,sijacki09,shen09}. While not
  directly comparable to these studies, as we focus on broad line quasars, 
  our results and conclusions are qualitatively consistent with 
  previous observational and theoretical work in that we find evidence
  for self-regulated SMBH growth, black hole downsizing, and BLQSO
  lifetimes of $t_{BL} \sim 150$ Myr.

  \section{SUMMARY}

  \label{s-summary}

  Our main results are:
  \begin{itemize}
  \item 
    We have, for the first time, obtained an estimate of the black
    hole mass function for broad-line quasars that self-consistently
    corrects for incompleteness and the statistical uncertainty in the
    mass estimates derived from the broad emission lines in a
    statistically rigorous manner. Our estimated BHMF was obtained
    using data from the SDSS DR3 quasar sample.
  \item
    The standard deviation in the statistical error of the broad line
    mass estimates is less than the commonly used $\sim 0.4$ dex
    within the range of luminosity and redshift probed in our
    analysis. This may be due to correlation between the error in the
    mass estimates and luminosity and/or redshift, or a dependence of
    the standard deviation of the error on luminosity and/or
    redshift. When we treat the standard deviation in the statistical error as a
    free parameter, we estimate that the the Mg II-based mass
    estimates scatter about the reverberation mapping estimates with
    an amplitude of $\approx 0.18$ dex, while the C IV-based estimate
    scatter with an amplitude of $\approx 0.13$ dex.
  \item
    We find evidence for cosmic downsizing among BLQSOs, where the
    number density of BLQSOs peaks at higher redshift with increasing
    black hole mass.
  \item 
    We find that the comoving mass density of SMBHs in BLQSOs
    peaks at $z \sim 2$. We use our estimate for the BHMF to
    place constraints on the duty cycle, $\delta(M_{BH},z)$, and
    lifetime, $t_{BL}$, for BLQSOs. The duty cycle at $z = 1$ is
    constrained to be $\delta \gtrsim 0.01$ at $M_{BH} \sim 10^{9}
    M_{\odot}$, falling to $\delta \gtrsim 10^{-5}$ at $M_{BH} \sim
    10^{10} M_{\odot}$. We estimate the lifetime of the BLQSO phase
    for SMBHs of $M_{BH} = 10^9 M_{\odot}$ at $z = 1$ to be $t_{BL} = 150 \pm
    15$ Myr. However, we will have underestimated the BLQSO lifetime if
    there is a population of $M_{BH} = 10^9 M_{\odot}$ SMBHs that
    never experience a BLQSO phase along our line of sight, or if the local number density of
    these black holes is significantly larger than the $z = 1$ number
    density of these black holes. We argue that our estimated BLQSO lifetime, in
    combination with the estimated Eddington ratio distribution, suggests
    that most of a SMBH's growth occurs when it is not seen as a BLQSO
    and accreting at a higher rate, and that BLQSO activity represents
    a short phase that most SMBHs go through, 
    consistent with self-regulated growth models.
  \item
    We estimate that the most massive SMBH that could be seen as a
    BLQSO is $M_{BH} \approx 3 \times 10^{10} M_{\odot}$. This SMBH
    would most likely be seen as a BLQSO at $z > 2$. While largely in
    agreement with previous work, we have for the 
    first time obtained statistically rigorous constraints on the value
    of $M_{BH}^{Max}$ and its redshift.
  \item 
    Assuming a constant bolometric correction of $C_{1350} = 4.3$
    \citep{vest09}, our inferred distribution of Eddington ratios
    peaks at $L / L_{Edd} \sim 
    0.05$ and has a dispersion of $\sim 0.4$ dex. Compared to previous
    work, our inferred Eddington ratio distribution is broader and
    shifted toward lower values of $L / L_{Edd}$, showing that
    previous estimated distributions of $L / L_{Edd}$ were
    significantly affected by incompleteness. We therefore provide
    evidence that most BLQSOs are not radiating at or near the
    Eddington limit, and that there is a large dispersion in Eddington
    ratio for BLQSOs. In addition, we also find that the number
    density of BLQSOs increases steeply toward lower values of $L / L_{Edd}$,
    consistent with models where the BLQSO phase occurs when the
    fuel supply is dwindling or halted.
  \item 
    The evolution of the cosmological SMBH mass density for BLQSOs
    tracks the evolution in the cosmological accretion rate 
    density of SMBHs estimated from variations of the \citet{soltan82}
    argument \citep{marconi04,merloni08,shank09}. This result, in
    combination with our estimated BLQSO lifetime and Eddington ratio
    distribution, are qualitatively consistent with models of
    self-regulated SMBH growth, with the BLQSO phase occuring at the end
    of the SMBHs obscured Eddington-limited growth
    \citep[e.g.,][]{stochacc,hopkins_long}.
  \end{itemize}

  \acknowledgements We thank Yue Shen, Priyamvada Natarajan, Martin
  Elvis, and Charles Steinhardt for
helpful discussions and comments on this paper, Mark Ammons and Aleks
Diamond-Stanic for helpful discussions, and the anonymous 
referee for a careful reading and comments that lead to improvement of
this work. BK acknowledges support by NASA through Hubble Fellowship
grants \#HF-01220.01 and \#HF-51243.01 awarded by the Space Telescope
Science Institute, which is operated by the Association of
Universities for Research in Astronomy, Inc., for NASA, under contract
NAS 5-26555. MV acknowledges financial support through grants
HST-AR-10691, HST-GO-10417, and HST-GO-10833 from NASA through the
Space Telescope Science Institute, which is operated by the
Association of Universities for Research in Astronomy, Inc., under
NASA contract NAS5-26555. XF acknowledges support by NSF grant
AST-0806861, a Packard Fellowship for Science and Engineering, a John
Simon Guggenheim Memorial Fellowship and the Max Planck Society. The
Dark Cosmology Centre is funded by the Danish National Research
Foundation.

\end{document}